\documentstyle[12pt]{article}
\def\AdSs5{$AdS_5$} 
\def\AdS5s5{$AdS_5 \times S^5$}

\def\RR{{$R\otimes R$}} 
\def\calE{{\cal E}}
\def\calZ{{\cal Z}} 
 
\def\calG{{\cal G}} 
\def\calN{{\cal N}} 
\def\Tr{\mbox{Tr}} 
\newcommand{\ie}{{\it i.e.~}}  
 
\newcommand{\s}{\sigma} 
 
\newcommand{\ap}{\alpha^{\prime}}  
\newcommand{\be}{\begin{equation}} 
\newcommand{\ee}{\end{equation}}  
\newcommand{\ba}{\begin{eqnarray}} 
\newcommand{\ea}{\end{eqnarray}} 
\newcommand{\ra}{\rangle}  
\newcommand{\la}{\langle} 
\begin{document}  
%%%%%%%%%%%%%%%%%%%%%%%%%%%%%%%%%%%%%%%%%%% 
%Feynman slash 
\newbox\SlashedBox  
\def\fs#1{\setbox\SlashedBox=\hbox{#1} 
\hbox to 0pt{\hbox to 1\wd\SlashedBox{\hfil/\hfil}\hss}{#1}} 
\def\hboxtosizeof#1#2{\setbox\SlashedBox=\hbox{#1} 
\hbox to 1\wd\SlashedBox{#2}} 
\def\littleFraction#1#2{\hbox{$#1\over#2$}} 
\def\ms#1{\setbox\SlashedBox=\hbox{$#1$}
\hbox to 0pt{\hbox to 1\wd\SlashedBox{\hfil/\hfil}\hss}#1}
\def\partialslash{\mathslashed{\partial}}
%%% 
\newcommand{\dd}{\raisebox{10pt}
           {\tiny$\scriptscriptstyle\longleftrightarrow$}\hspace{-11.7pt}}
% 
%%% 
\def\I {\mbox{1}\hspace{-1.1mm}\mbox{I}} 
\def\R {\mbox{I}\!\mbox{R}}
\def\C {\hspace{1.35mm}\mbox{\rule{0.6pt}{8pt}}\hspace{-1.35mm}\mbox{C}}
\def\sC {\hspace{0.75mm}\mbox{\rule{0.3pt}{4pt}}\hspace{-0.75mm}\mbox{C}} 
\def\Z {\mbox{\sf{Z}}\hspace{-1.6mm}\mbox{\sf{Z}}\hspace{0.4mm}}
%
%%%%%%%%%%%%%%%%%%%%%%%%%%%%%%%%%%%%%%%%%%%%  
%Without pictures use this macro  
\def\pct#1{(see Fig. #1.)}  
 
%With  pictures use this macro  
%\def\pct#1{\input epsf\centerline{\epsfbox{#1.eps}}} 
 
%%%%%%%%%%%%%%%%%%%%%%%% FRONT PAGE  
%%%%%%%%%%%%%%%%%%%%%%%%%%%%%%%%%%%%% 
 
\begin{titlepage} {\hfill ROM2F-98-25;~ DAMTP-98-69;~ hep-th/9807033} 
 
\begin{center}   
\vspace{0.8cm} {\centerline {\large \bf Instantons in supersymmetric 
Yang--Mills 
and  }} 
\vskip 0.25cm  
{\centerline {\large \bf D-instantons in IIB superstring theory}} 
\vskip 0.8cm 
{\centerline{  Massimo Bianchi$^a$, Michael B. Green$^b$, Stefano 
Kovacs$^a$}}
{\centerline { and Giancarlo Rossi$^a$ }}  
 
\vspace{0.6cm} 
 
{\sl $\ ^a$ Dipartimento di Fisica, \ \ Universit{\`a} di Roma \  
``Tor 
Vergata'' \\ I.N.F.N.\ - \ Sezione di Roma \ ``Tor Vergata'', \\ Via 
della  
Ricerca  Scientifica, 1 \\ 00173 \ Roma, \ \ ITALY} 
\vskip 0.3cm 
\centerline{\sl $\ ^b$ Department of Applied Mathematics and  
Theoretical 
Physics,} 
\centerline{\sl Silver Street, Cambridge CB3 9EW, UK} 
 \vspace{1.0cm} 
 \end{center} 
\centerline{ABSTRACT} 
 
\vspace{0.3cm} 
 
The one-instanton contributions to various 
correlation functions of superconformal currents in four-dimensional 
${\cal N}=4$ supersymmetric  
$SU(2)$ Yang--Mills theory are evaluated to the lowest order in 
perturbation theory.   
Expressions of the same form are obtained from the leading effects 
of a single D-instanton extracted from the IIB 
superstring effective  action  around the \AdS5s5\ background.  This 
is in 
line with the suggested $AdS$/Yang--Mills correspondence. 
The relation between 
Yang--Mills instantons and D-instantons is further confirmed by the 
explicit form of the classical D-instanton solution in the \AdS5s5\ 
background and its associated supermultiplet of zero modes.
Speculations  are made concerning instanton effects in  the 
large-$N_{c}$ limit of the $SU(N_{c})$ Yang--Mills theory.    
  
\vspace{0.6cm}

\vspace{1cm}

%\underline{\today}

\end{titlepage}

\section{Introduction} 
 
Four-dimensional ${\cal N}=4$  supersymmetric Yang--Mills theory \cite{bssgos}  
is a very special quantum field theory. It is the original example of a theory 
possessing exact electromagnetic duality 
\cite{olivemont,osborna,olivewit} which is connected to the fact that 
it contains an infinite set of stable dyonic BPS states \cite{sen} and 
also has a vanishing renormalization group $\beta$ function \cite{fz}.  
Whereas the abelian theory is free,  the nature of the non-abelian
theory depends on whether scalar fields have vacuum expectation
values.  In the Coulomb phase,  reached by 
giving vacuum expectation values  (vev's) to the scalars in the 
Cartan  
subalgebra,  the  two-derivative Wilsonian effective action  is
believed not to be renormalized either in perturbation theory or    
by non-perturbative effects.  The superconformal invariance of the
theory is,  
however, broken in  this phase.  The phase in which all scalar fields 
have 
vanishing   vev's  is expected to describe a highly  nontrivial 
superconformal 
field theory.  Although it is very difficult to understand the nature 
of this 
phase from direct perturbative calculations, according to the recent 
flurry of 
work 
\cite{gubskleb,maldacena,gkp,wittone} certain properties  of the 
theory should 
be understandable in terms of type IIB superstring theory 
compactified on 
\AdS5s5, where the Yang--Mills theory is located on the 
four-dimensional 
boundary of the five-dimensional anti de-Sitter space.   
 
According to the proposal made by Maldacena in \cite{maldacena} 
properties of 
$SU(N_c)$ 
$\calN=4$ Yang--Mills theory in the  large-$N_c$ limit may be  
determined by 
semi-classical approximations to  the superstring theory.  In this 
limit the 
boundary Yang--Mills theory is interpreted as the world-volume theory 
for a 
large-$N_c$ collection of coincident D3-branes.  The \AdS5s5\ 
geometry is the 
near-horizon description of the classical D3-brane  solution 
\cite{duff3} 
which is a source of non-vanishing self-dual Ramond--Ramond (\RR) 
five-form field strength, $F_5  = *F_5$ \cite{cjp}. 
In fact, the D3-brane solution 
  plays the r\^ole of an  interpolating soliton between two 
maximally supersymmetric  configurations (with 
$16+16$ supercharges) ---  flat ten-dimensional Minkowski space at  
infinity 
and \AdS5s5\  at  the horizon. The extra $16$ 
Killing spinors at the horizon are in  one-to-one  correspondence 
with the 
special supersymmetry transformations of the  boundary theory 
\cite{fgk}.  
The $SO(6)$ 
isometry of the $S^5$ factor  corresponds to the gauging of the 
$SU(4)$ 
R-symmetry group while the  $SO(4,2)$  isometry of 
$AdS_5$ coincides with the conformal group of the boundary theory at  
the horizon.  
 
In this picture the boundary values of the  fields of the   bulk 
superstring 
theory   compactified on \AdS5s5\   are sources that couple to 
gauge-invariant  
operators of the four-dimensional 
 boundary $\calN=4$ supersymmetric  Yang--Mills theory. The 
lowest Kaluza--Klein modes of the graviton   supermultiplet couple to 
the 
superconformal multiplet of Yang--Mills  currents. These fields 
and currents will be reviewed in  more detail in section 2. 
More generally, all the  Kaluza-Klein excitations of the bulk 
supergravity theory  can be put in one-to-one  correspondence with  
gauge-singlet composite Yang--Mills operators 
\cite{gkp,wittone,adf,ff}.
 
According to this idea the effective action of type IIB supergravity, 
evaluated 
on a solution of the equations of motion   with prescribed boundary 
conditions, is equated with the generating  functional of connected  
gauge-invariant correlation functions in 
the  Yang--Mills   theory.    The  parameters of the $\calN=4$ 
Yang--Mills 
theory and the IIB  superstring on \AdS5s5\ are related by 
\begin{equation} 
\label{dict} g_s = {g_{_{YM}}^2 \over {4\pi}} \:, \qquad
2\pi\tilde  C^{(0)} =  \theta_{_{YM}}\:, \qquad {L^2 \over 
\ap} = 
\sqrt{g^2_{_{YM}}N_c} \: , 
\end{equation}   
where $g_s=e^{\tilde \phi}$ is the string coupling ($\tilde 
\phi$ is the constant  dilaton), 
$\tilde C^{(0)}$ is the constant  \RR\ axionic background,
$g_{_{YM}}$ is
the   Yang--Mills coupling, $\theta_{_{YM}}$ is the 
vacuum angle and $L$ is the  radius of both the \AdSs5\ and  
$S^5$ 
factors of the bulk background. 
The complex Yang--Mills coupling
 is therefore identified
with the constant boundary value of the complex scalar field 
of the IIB superstring,
\be
\label{idenscal}
{\theta_{_{YM}}\over 2\pi}  +  {4\pi i\over g_{_{YM}}^2} =
\tilde{C}^{(0)} +  {i\over g_s}  .
\ee
Most of the tests of this conjecture 
have so 
far amounted to the computation of two-point and three-point 
correlations 
of currents 
\cite{twoandthree} 
based  on  the semiclassical approximation to the bulk supergravity 
($g_s <<  
1$) which is valid at length scales much larger than the string 
scale  or, 
equivalently, in the limit $\ap/L^2 << 1$.  This is the limit $N_c 
\to \infty$ 
and $g_{_{YM}} \to 0$ with  
$g_{_{YM}}^2 N_c >> 1$.   This can also be viewed as the large-$N_c$ 
limit  
introduced by 't Hooft \cite{thooft} with the coupling  $\hat g^2 =   
g_{_{YM}}^2 
N_c$ fixed at a large value. 
 
The explicit connection between the bulk theory and the boundary  
theory can be  expressed symbolically as \cite{wittone,gkp,ffz}
\be 
\exp(-S_{_{IIB}} [\Phi_{m}(J)]) = \int DA \, \exp(-S_{_{YM}}[A] + 
{\cal O}_{\Delta} [A] J), 
\label{main} 
\ee 
where $S_{_{IIB}}$ is the effective action  of the IIB superstring or 
its low 
energy supergravity limit which  is evaluated  in terms of the \lq 
massless' 
supergravity fields  and their Kaluza--Klein  
descendents, that we have generically indicated with 
$\Phi(z;\omega)$, where $\omega$ are the coordinates on 
$S^5$ and  
$z^M\equiv (x^\mu,\rho)$ ($M=0,1,2,3,5$ and $\mu=0,1,2,3$) are the 
$AdS_5$ 
coordinates  ($\rho \equiv z_5$ is  
the coordinate transverse to the boundary).  The notation in 
(\ref{main}) 
indicates that the action depends on the boundary values,    $J(x)$, 
of the 
bulk fields.  The 
fluctuating boundary $\calN=4$ supersymmetric  Yang--Mills  fields 
are denoted 
by $A$  and ${\cal O}(A)$ in (\ref{main})  
is the set of gauge-invariant  composite operators to which $J$ 
couples.   The recipe for  computing correlations involves the \lq 
bulk-to-boundary'  
Green functions which are defined as specific normalized limits of
bulk-to-bulk Green functions  \cite{gkp,wittone,freedman} when one
point is taken to the $AdS$ boundary.  
The precise forms of these propagators 
depend on the spin and mass  
of the field. For example, the normalized bulk-to-boundary   
Green function for a dimension  $\Delta$ scalar field
is given by 
\be
\label{gdef}
G_{\Delta}(x,\rho,\omega; x',0,\omega') =  c_{_{\Delta}} K_{\Delta} 
(x^\mu, 
\rho; x'^{\mu},0),
\ee
which is independent of $\omega$ and 
where  $c_{_\Delta} = \Gamma(\Delta)/(\pi^2 \Gamma(\Delta - 2))$ and 
\be
 K_{\Delta} (x^\mu, \rho; x'^{\mu},0) =  
{\rho^{\Delta} \over  (\rho^2 
+ (x-x')^2)^{\Delta}}. 
\label{greenfun} 
\ee   
The expression (\ref{gdef}) is appropriate for 
an `S-wave' process in which there are no excitations 
in the directions of the five-sphere, $S^5$.  
In terms of $K_{\Delta}$ the bulk field 
\be
\Phi_m (z;J) = c_{_{\Delta}}\int d^4x'  
K_{\Delta} (x,\rho; x',0) J_{\Delta}(x') 
\label{btob}
\ee
satisfies the boundary condition as $\rho \to 0$,
\be
\Phi_m (x,\rho;J) \approx \rho^{4-\Delta} J_{\Delta}(x)  
\label{boundcond}
\ee
since $\rho^{\Delta-4}K_{\Delta}$ reduces to a $\delta$-function on the 
boundary.  The conformal dimension 
of the operator is related to the $AdS$ mass of the corresponding 
bulk 
field by $(mL)^2 = \Delta(\Delta -4)$,  
so that $\Delta_{\pm} = 2 \pm \sqrt{4+(mL)^2}$ and only the positive 
branch,  
$\Delta = \Delta_+$, is relevant for the lowest-\lq mass' 
supergravity  
multiplet.  In the case of a massless scalar field ($\Delta_+ =4$) 
the propagator reduces to $\delta^{(4)}(x^\mu - x'^{\mu})$ in the 
limit 
$\rho \to 0$.  
 
For our considerations it will prove crucial in the following that the
expression  (\ref{greenfun}) in the case $\Delta_+ =4$ has exactly 
the same form  as the 
contribution of a  Yang--Mills instanton to $\Tr  (F^-_{\mu\nu})^2$ 
(where 
$F^-_{\mu\nu}$ is the non-abelian self-dual field strength) when 
the  fifth  
coordinate $\rho$ is identified with the instanton scale.  At the 
same time, we  
will see that in this case (\ref{greenfun}) has  precisely the same 
form as the  five-dimensional 
profile of a  D-instanton centered on the point $z^M$  in  
\AdSs5\  and evaluated at the boundary point $(x^{\prime
\mu},0)$.   
This is a key observation in  identifying   D-instanton  effects of 
the bulk theory with Yang--Mills instanton effects in  the boundary 
theory.   It is related to the fact that the moduli space of a
Yang--Mills instanton has an $AdS_5$ factor.
  
Two-point and three-point correlation functions of superconformal 
currents
 are  not renormalized from their  free-field  values  due to  the 
$\calN=4$ superconformal invariance so   
they  do not get interesting interaction corrections 
\cite{freedman,afgj,howest,sei}. 
 However,  
higher-point correlation functions do receive nontrivial interaction 
corrections.  Here we will be concerned with calculations of 
processes 
in which  
the one-instanton contributions can be evaluated exactly  to lowest 
order in perturbation theory. The 
literature on this subject includes examples in which all scalar 
field vacuum expectation 
values are zero ($v=0$) \cite{akmrv} as well as those in which the 
scalar 
fields in the Cartan subalgebra 
have non-vanishing vacuum expectation values  ($v\ne 0$) 
\cite{shifman}. In the latter case   
the theory is in an abelian Coulomb phase where the dynamics is 
rather trivial.   
The superconformal theory of interest here has $v=0$ 
and cannot be obtained  
simply as the limit of the theory with $v\neq 0$, 
since all the  higher  
derivative terms that naturally have inverse powers of $v$ 
become singular.   
Although there are few explicit computations of  nontrivial 
instanton effects  in the literature in the superconformal ${\cal 
N}=4$ 
case  where $v=0$ (see, however \cite{vafawitten,bfmr}),  
we will see  that there is no impediment
to  using well known methods \cite{akmrv}  
to perform such calculations.  

In section 3  we will evaluate the contributions due to a  single  
 $SU(2)$ instanton to specific correlation functions in the 
 Yang--Mills theory.  
One class of correlation functions that we will discuss  is  
the product of four bilinear operators in the  current 
supermultiplet,  
$\langle {\cal W}_{(2)}(x_1) \dots {\cal 
W}_{(2)}(x_4) \rangle$.  For example, we will consider the correlator 
of the 
components, ${\cal Q}$, of ${\cal W}_{(2)}$ that are bilinears in the 
scalar fields $\varphi$  
belonging to the  six-dimensional representation of the $SU(4)$ 
R-symmetry. This correlation function receives (at 
$\theta_{_{YM}}=0$) 
identical contributions from instantons and anti-instantons.
We will also consider the correlation 
function of sixteen fermionic superconformal current operators, 
$\langle {\hat \Lambda}(x_1) \dots {\hat \Lambda}(x_{16})\rangle$, 
where 
$  {\hat \Lambda}  $ is the product of $F^-$ (the self-dual 
field strength) 
and $\lambda$ (the spin-1/2 gaugino in the {\bf 4} of  
$SU(4)$) and  the  correlation function of eight 
gaugino bilinears,  
$\langle {\cal E}(x_1) \dots {\cal E}(x_8) \rangle$. 
In contrast to the four-point function of ${\cal Q}$'s, these 
correlators receive non vanishing 
contributions only from instantons and not from anti-instantons 
(to leading order in the
 Yang--Mills coupling). The common feature of all the
correlation functions under consideration, as well as many others  
that  are  
related  by supersymmetry, is that they provide precisely the sixteen
 fermionic  
zero modes that are needed to give a nonzero result in the 
(anti-)instanton 
background.   

As noted in \cite{banksgreen} those effects 
 that are seen from the Yang--Mills perspective as instanton effects 
 will be seen from the bulk point of view as effects due to 
 D-instantons.  Explicit   D-instanton effects can be extracted from 
certain  
terms in the IIB 
 effective action that (for fixed radius $L$) are of order 
 $(\ap)^3$   
%$\sim (g^2_{_{YM}} N_c)^{-3/2}$ 
relative to the leading 
 Einstein--Hilbert term and have been derived in 
 \cite{greengut,greenvanhove}.  One example of such a term has the 
 form  (in the Einstein frame)  
$(\ap)^{-1} \int d^{10}X\sqrt G e^{-\phi/2} 
f_4(\tau,{\overline \tau}) {\cal R}^4$ 
where 
 $\tau$ is the   
complex scalar field,  
 $\tau = C^{(0)} + i e^{-\phi}$,  and ${\cal R}^4$ 
denotes a 
 particular contraction of four ten-dimensional Riemann curvature 
 tensors.  The function $f_4(\tau,{\overline \tau})$ is a 
nonholomorphic 
 Eisenstein series that may be expanded for small string coupling 
 ($\tau_2 = e^{-\phi} \to \infty$) in an infinite series of 
 D-instanton terms, in addition to a tree-level and one-loop 
 contribution \cite{greengut}.   This D-instanton contribution  will 
be reviewed in section 4 as will the analogous contribution to the 
sixteen-dilatino term, 
 $(\ap)^{-1} \int d^{10}X \sqrt G e^{-\phi/2} f_{16} (\tau,{\overline 
\tau})\Lambda^{16}$ 
(where $f_{16}$ transforms with specific 
 holomorphic and antiholomorphic weights),  which was obtained in  
\cite{greengutkwon}.   The many other related terms of the same 
dimension 
in the IIB effective action that are of the same order  in $\alpha'$ 
can be 
obtained in similar fashion. 

The leading effects of a single D-instanton that are extracted in 
this manner will be compared in section 5  with the corresponding 
$\calN=4$ 
 Yang--Mills supercurrent correlators that were  considered in 
section 3. 
The correspondence between these expressions 
will be demonstrated most completely for the sixteen-fermion 
amplitude.    
We will see that the form of the   
leading charge-one  D-instanton contribution to 
 the $\Lambda^{16}$  amplitude  
matches with  the expression for the sixteen-${\hat \Lambda}$
correlator in $\calN=4$ $SU(2)$ 
Yang--Mills theory calculated in section 3.
 Given this
 correspondence, supersymmetry guarantees that   all 
the related correlators
 must also agree with their $AdS$ counterparts.  Although we have not 
completed a detailed comparison for all such correlation functions we 
will indicate in outline how this correspondence should work in the 
cases described in section 3.  

In order to further clarify the correspondence between Yang--Mills 
instantons and D-instantons, 
in section 6 we will present the  classical D-instanton solution of  
the type IIB supergravity equations in the 
\AdS5s5\ background.  The D-instanton is a solution to the 
ten-dimensional 
 Euclidean field equations that breaks half the supersymmetries and 
in which the complex scalar IIB field, $\tau$, has non-trivial 
spatial 
dependence.  
The solution is particularly simple to establish by using the fact 
that 
\AdS5s5\ is conformally flat, which relates the solution to the 
flat-space solution of \cite{ggp}. Although the D-instanton 
solution does not affect the \AdS5s5\ geometry in the Einstein frame 
it generates a
wormhole in the string frame which leads to a rather intriguing
modification of the geometry in the limit of large instanton number 
($K\to \infty$).

The superpartners of the 
D-instanton in the gravity supermultiplet are obtained  by acting on 
this solution  with the broken supersymmetries
and it is striking that the 
D-instanton induced profiles of these fields 
match with those  of the corresponding Yang--Mills
 supercurrents.  This reflects the fact that the moduli space of a 
$SU(2)$ Yang--Mills instanton is $AdS_5$ (related comments appear  
in  \cite{wittennew}).
We will also verify in section 6  that 
the leading
D-instanton contribution to the $\Lambda^{16}$ amplitude  can be 
obtained  by  semiclassical quantization around the classical D-instanton 
background. 

It is encouraging to find this level of agreement between the 
instanton terms in the IIB superstring and the Yang--Mills theory.  
Of course, it is difficult to assess without further calculations 
whether the agreement is qualitative or more precise.  The
 Yang--Mills calculation  is based on perturbation theory  and with 
the gauge group $SU(2)$  
while agreement with the $\alpha'$ expansion of the  IIB theory 
is only expected in  the limit of large $N_c$ and 
$g^2_{_{YM}} N_c$.   
We only  consider the leading term in the 
 semiclassical expansion around a single-instanton background while  
there are known to be an infinite number of 
 perturbative corrections even to the one D-instanton expression.
For $N_c >2$ 
there are additional issues  related to the presence of extra zero 
 modes in the  
semi-classical  approximation to the Yang--Mills theory.  
We will discuss this and other aspects of the generalization  
to $SU(N_c)$ in  the  concluding section.

\section{Fields  and Currents in ${\cal N}=4$ SYM } 
 
The field content of the maximally supersymmetric (${\cal N}=4$)  
four-dimensional Yang--Mills \cite{bssgos,fz,foursusy} theory  is unique 
apart from the choice  of the 
gauge group.  The theory is  classically invariant under 
superconformal 
transformations as well as  under global $SU(4)$ transformations, 
which form the R-symmetry 
group of  automorphisms of the ${\cal N}=4$ supersymmetry algebra.  
 The field content  consists of six real scalars,  four Weyl 
spinors and one vector which are all in the adjoint  representation 
of  the 
gauge group. More precisely, the scalars $\varphi^{AB} = - 
\varphi^{BA}$ (with 
${\overline 
\varphi}_{AB}=\frac{1}{2}\varepsilon_{ABCD}\varphi^{CD}$)  are in 
the {\bf 6} of the 
$SU(4)$ R-symmetry group\footnote{The superscripts $A,B=1,\dots,4$ 
label {\bf 
4}'s of $SU(4)$ while  subscripts label the {\bf 4}$^*$'s.}.  In the 
following 
we will also use  
the notation $\varphi^{i}=\frac{1}{2} {\overline{t}^{i}}_{AB} 
\varphi^{AB}$ ($i=1,\ldots,6$) with  $\varphi^{i} =  
{\varphi^{i}}^{*}$ and $\bar t^i_{\ AB}$ are  
Clebsch--Gordan coefficients that  couple two 
{\bf 4}'s to a {\bf 6} (these are six-dimensional generalizations of 
the four-dimensional
${\sigma^{\mu}}_{\alpha {\dot \alpha}}$ matrices).   The spinors  
$\lambda^{A},~{\overline \lambda}_{B}$ transform  as {\bf 4}  and 
{\bf 4}$^{*}$, respectively, and the vector $A_{\mu}$ is a  singlet 
of $SU(4)$. 
The classical Minkowskian action is given in terms of these component 
fields by 
\begin{eqnarray} 
	S &=& \int d^{4}x \: \Tr \Big\{ 
    (D_{\mu}\varphi^{AB})(D^{\mu}{\overline \varphi}_{AB}) 
	-\frac{1}{2}i({\lambda^{\alpha}}^{A} 
	{\dd \ms{D}}_{\alpha{\dot \alpha}} 
	{{\overline{\lambda}}^{{\dot \alpha}}}_{A}) - 
	\frac{1}{4}F_{\mu \nu}F^{\mu \nu} \nonumber \\ 
 	&& \hspace{-1.5cm} - g_{_{YM}} {\lambda^{\alpha}}^{A}
 	[{\lambda_{\alpha}}^{B}, {\overline{\varphi}}_{AB}] 
	-g_{_{YM}} {\overline{\lambda}}_{{\dot \alpha}A}[{{\overline 
    {\lambda}} ^{{\dot \alpha}}}_{B}, \varphi^{AB}] + 2g^{2}_{_{YM}}  
	[\varphi^{AB},\varphi^{CD}][{\overline{\varphi}}_{AB}, 
	{\overline{\varphi}}_{CD}]\Big\}~~~ 
	\label{action} 
\end{eqnarray} where $\Tr$ denotes a trace over the $SU(N_c)$ colour 
indices and $D_{\mu}^{ab} = \delta^{ab} \partial_{\mu} + i g_{_{YM}} 
f^{abc} A_{\mu c}$ is the covariant derivative in the adjoint 
representation. 
 
Although there is no off-shell formulation of ${\cal N}=4$  
Yang--Mills theory the on-shell states may be  packaged into  a 
superfield 
$W^{AB}$ that is a function of sixteen Grassmann  
coordinates, $\theta^A_\alpha$ and ${\overline \theta}^{\dot 
\alpha}_{A}$,  
  satisfying  
\cite{townsenda} the reality  condition, 
\be  
{\overline W}_{A B} ={1\over 2} \varepsilon_{A B C D} W^{C D},  
\label{realw} 
\ee  
together with the constraint, 
\be
 {\cal D}_{\alpha}^{A}  W^{BC} = {\cal D}_{\alpha}^{[A} W^{BC]}, 
\label{confag} 
\ee  
where ${\cal D}_\alpha$ is the super--covariant derivative.  The 
latter 
equation  constrains the component fields to satisfy  the equations 
of motion. It follows that  the supersymmetry transformations on the
component  
fields  with parameters ${\eta_{\alpha}}^{A}$ and 
${\overline\eta^{\dot\alpha}}_{A}$ are given by,
\begin{eqnarray} 
	\delta \varphi^{AB} &=& \frac{1}{2} ({\lambda^{\alpha}}^{A}  
    {\eta_{\alpha}}^{B}- 
	{\lambda^{\alpha}}^{B} {\eta_{\alpha}}^{A}) + \frac{1}{2}  
        \varepsilon^{ABCD} 
	{{\overline{\eta}}_{\dot \alpha}}_{C}  
	{\overline \lambda}^{\dot \alpha}_{D} \nonumber \\ 
	\delta {\lambda_{\alpha}}^{A} &=& -\frac{1}{2} F^-_{\mu \nu}  
	{{\sigma^{\mu \nu}}_{\alpha}}^{\beta} {\eta_{\beta}}^{B} + 4i 
        (\ms{D}_{\alpha {\dot \alpha}} \varphi^{AB}) 
        {\overline{\eta}}^{{\dot \alpha}}_{B} - 
	8g_{_{YM}}[{\overline{\varphi}}_{BC}, \varphi^{CA}] 
        {\eta_{\alpha}}^{B}  \nonumber \\ 
	\delta A_{\mu} &=& -i {\lambda^{\alpha}}^{A}  
	{\sigma^{\mu}}_{\alpha{\dot \alpha}} 
	{{\overline \eta}^{\dot \alpha}}_{A} 
	-i {\eta^{\alpha}}^{A} {\sigma^{\mu}}_{\alpha {\dot \alpha}} 
	{{\overline \lambda}^{\dot \alpha}}_{A} . 
	\label{transs} 
\end{eqnarray} 
 
The classical action~(\ref{action}) is superconformally invariant 
\cite{foursusy} and  this 
property is believed to be preserved at the quantum level thanks to 
the 
exact vanishing of the $\beta$-function \cite{fz}. The Noether  currents 
associated with  
the superconformal transformations,  together with  those 
corresponding to 
chiral $SU(4)$ transformations, constitute a  supermultiplet whose 
components are  bilinears of  $W^i \equiv {1\over 2}t_{AB}^i W^{AB}$  
%($i =1, \dots, 6$), 
\be 
  {\cal W}_{(2)}^{ij} = \Tr(W^i W^j - {\delta^{ij} \over 6} W_k W^k) 
\, . 
  \label{supcur} 
\ee  
The components of this  
current 
superfield are given for the abelian case by \cite{dewit}, 
\begin{eqnarray} 
	{\cal T}^{\mu \nu} &=& \frac{1}{2} [\delta^{\mu \nu} 
	(F^{-}_{\rho \sigma})^{2}-4{{F^{-}}^{\mu}}_{ \rho} {F^{-}}^{\nu  
    \rho} +  
	\mbox{ h.c.}] - \frac{1}{2} \lambda^{\alpha A} 
	{\sigma^{( \mu}}_{\alpha {\dot \alpha}} 
	\dd \partial^{\nu )}{\overline{\lambda}}^{{\dot \alpha}}_{A} 
	\nonumber \\ 
	&+& \delta^{\mu \nu} (\partial_{\rho} 
	{\overline{\varphi}}_{AB})(\partial^{\rho} \varphi^{AB})  
	-2(\partial^{\mu}{\overline{\varphi}}_{AB}) 
	(\partial^{\nu} \varphi^{AB})  \nonumber \\ 
	&-& \frac{1}{3} (\delta^{\mu \nu} \Box 
	- \partial^{\mu }\partial^{\nu})({\overline{\varphi}}_{AB}  
	\varphi^{AB})  \nonumber \\ 
	{\Sigma^{\mu}}_{\alpha A} &=& -\sigma^{\kappa \nu} 
    F^{-}_{\kappa \nu}  
	{\sigma^{\mu}}_{\alpha {\dot \alpha}}  
	{{\overline \lambda}^{{\dot \alpha}}}_{A} + 
	2i {\overline \varphi}_{AB} \dd \partial^{\mu} 
	{\lambda_{\alpha}}^{B} + \frac{4}{3}i 
	{{\sigma^{\mu \nu}}_{\alpha}}^{\beta}  
	\partial_{\nu}({\overline \varphi}_{AB} {\lambda_{\beta}}^{B})  
    \nonumber \\ 
	{{{\cal J}^{\mu}}_{A}}^{B} &=& {\overline{\varphi}}_{AC} 
    \dd \partial^{\mu}  
	\varphi^{CB} +  {\overline \lambda}_{{\dot \alpha} A} 
	{\overline \sigma}^{\mu{\dot \alpha}\alpha}  
	{{\lambda}_{\alpha}}^{B} 
	-\frac{1}{4} {\delta_{A}}^{B} \lambda^{\alpha C} 
	{\sigma^{\mu}}_{\alpha{\dot \alpha}}  
	{{\overline{\lambda}}^{\dot \alpha}}_{C} \nonumber \\ 
	{\cal C} &=& ({F^{-}}_{\mu \nu})^{2}  \nonumber \\ 
	{{\hat \Lambda}_{\alpha}}^{A} &=& 
	-{{\sigma^{\mu \nu}}_{\alpha}}^{\beta}  
	{F^{-}}_{\mu \nu} {\lambda_{\beta}}^{A}  \nonumber \\ 
	{\cal E}^{AB} &=& {\lambda^{\alpha}}^{A}  
	{\lambda_{\alpha}}^{B} \nonumber \\ 
	{{\cal B}_{\mu \nu}}^{AB} &=& \lambda^{\alpha A}  
	\sigma_{\mu \nu\alpha}{}^{\beta} 
	{\lambda_{\beta }}^{B} + 2i {\varphi}^{AB} {F^{-}}_{\mu \nu} 
     \nonumber \\ 
	{\hat \chi}_{\alpha\, AB}^C &=& \frac{1}{2}  
     \varepsilon_{ABDE}(\varphi^{DE}  
	{\lambda_{\alpha}}^{C}+\varphi^{CE} {\lambda_{\alpha}}^{D})  
     \nonumber \\ 
	{{\cal Q}^{AB}}_{CD} &=& \varphi^{AB} {\overline{\varphi}}_{CD} -  
    \frac{1}{12}{\delta^{A}}_{[C} {\delta^{B}}_{D]} \varphi^{EF}  
	{\overline{\varphi}}_{EF} \;  .
\label{currentdef} 
\end{eqnarray}  
In the non-abelian case there is 
a trace over the $SU(N_c)$ colour indices and  there are additional 
terms  that enter, for example, into the covariantization of all the 
derivatives.   In (\ref{currentdef}), ${\cal T}_{\mu\nu}$  
is  the  energy-momentum tensor, ${\Sigma^{\mu}}_{\alpha A}$ are the 
supersymmetry  currents and  ${{{\cal J}^{\mu}}_{A}}^{B}$ the  
$SU(4)$ R-symmetry currents.   
The remaining components are obtained by supersymmetry using the 
equations of motion. They 
consist of three  scalar components  (${\cal C}$, ${\cal E}^{(AB)}$, 
${\cal Q}^{ij}$), two fermionic spin-1/2 components  
(${{\hat \chi}^C}_{AB}$ and ${\hat \Lambda^A}$) 
and  one antisymmetric tensor (${{\cal B}_{\mu\nu}}^{[AB]}$). 
 
It is often useful to decompose the ${\cal N}=4$ multiplet in  terms 
of either ${\cal N}=1$ or ${\cal N}=2$ multiplets.  
In the ${\cal N}=1$ description a $SU(3) \times U(1)$ subgroup of 
the original  
$SU(4)$ is  manifest and the representations decompose according to 
{\bf 6} $\rightarrow$ {\bf 3}+{\bf 3}$^{*}$,  {\bf 4} 
$\rightarrow$ {\bf 3}+{\bf 1}.  Thus, the  ${\cal N}=1$ decomposition 
of the ${\cal N}=4$  field 
supermultiplet consists of one $\calN=1$ vector supermultiplet, $V$, 
and three 
$\calN=1$ chiral  supermultiplets, 
$\Phi^{I}$ ($I=1,2,3$). The vector, $A_{\mu}$, and  the Weyl 
spinor, 
$\lambda_{\alpha}$, in $V$ are $SU(3)$ singlets  whereas the 
complex scalars  
$\phi^{I}$ and the Weyl spinors $\psi^{I}$ belong to chiral 
multiplets 
in the {\bf 3}.  The Yukawa couplings and scalar self-interactions 
are generated by the manifestly $SU(3)$ invariant superpotential 
\be 
\mbox{W} = {1\over 3!} g_{_{YM}} \varepsilon_{_{IJK}} 
f_{abc} \Phi^{aI} \Phi^{bJ} \Phi^{cK} \, .
\label{superpot}
\ee 
In the ${\cal N}=2$ description  the manifest  global symmetry is 
$SU(2)_{{\cal V}}\times SU(2)_{{\cal H}} \times U(1)$ and the ${\cal 
N}=4$ field supermultiplet   
decomposes into a $\calN=2$ vector multiplet ${\cal V}$ and a 
hypermultiplet,  
${\cal H}$.  
We will denote  the representations of  $SU(2)_{\cal V} \times 
SU(2)_{\cal H} 
\times U(1)$  by  ({\bf r}$_{{\cal V}}$, {\bf r}$_{{\cal 
H}}$)$_{q}$  with the subscript $q$ referring to  the  
$U(1)$ charge  and will make use of the following notation for the  
component 
fields, 
 \begin{eqnarray} 
 	{\cal V} ~~~ && \rightarrow ~~~ \lambda^{u} \in {({\bf 2}, 
        {\bf  1})}_{+1}, 
 	~~~\varphi \in {({\bf 1}, {\bf 1})}_{+2}, ~~~A_\mu \in {({\bf 1}, 
        {\bf 1})}_{0} \nonumber \\ 
 	{\cal H} ~~~ && \rightarrow ~~~ \psi^{{\dot u}} \in {({\bf 1},  
 	{\bf 2})}_{-1}, ~~~ q_{u{\dot u}} \in {({\bf 2}, {\bf 2})}_{0} ~~,  
 	\label{entwo}
 \end{eqnarray}  where $u,{\dot u} = 1,2$. The scalar 
component 
${{\cal Q}^{AB}}_{CD}$ of  the  
${\cal N}=4$ current decomposes in the following way in terms of  
${\cal N}=2$ fields, 
\begin{eqnarray} 
	{\cal Q}^{S 0} &= &q^{S}\varphi\qquad {{\cal Q}^{S 0}}^{\dagger} = 
q^{S} {\overline \varphi}  \qquad {\cal Q}^{S T} = 
q^{S}q^{T} - \mbox{ trace} 
	  \nonumber \\ 
	{\cal Q}_{(+)} & =& \varphi^{2} \qquad\quad
	{\cal Q}_{(-)} = {\overline \varphi}^{2}\qquad\quad  
	{\cal Q}_{(0)}  = {\overline \varphi} \varphi - \mbox{ trace}~~,  
\label{dqfields} 
\end{eqnarray} where $q_{u {\dot u}} = q_{_S} {\sigma^{^{S}}}_{u 
{\dot u}}$ 
and $\sigma^{^{S}}=(\I,\vec\sigma)$, with $\vec\s$ the standard Pauli 
matrices (and $S,T = 1,\dots,4$ are $SU(2)_{\cal V} \times 
SU(2)_{\cal H}$ 
vector indices). 
 
In calculating the correlation functions in the next section it will  
be 
important to understand the systematics of the fermion zero modes  
associated 
with the Yang--Mills instanton.  As we will see these can  either  be 
determined directly from the supersymmetry  transformations 
(\ref{transs})  
or by making use of the  explicit  Yukawa couplings which are given 
in 
$\calN=2$ notation by, 
\begin{eqnarray} 
	{\cal L}^{({\cal N}=2)}_{Y} = {\sqrt{2}\over 2} g_{_{YM}} f_{abc} 
&& \hspace{-0.8cm} \left\{ 
	q^a_{_S}\left({\lambda^{b\alpha}}^{u}  
	{\sigma^{^{S}}}_{u {\dot u}} {\psi_{\alpha}}^{{c\dot u}} + 
	{{\overline{\psi}}^b_{{\dot \alpha}}}_{{\dot u}}  
	{{\overline{\sigma}}^{^{S}}}^{{\dot u} u}  
	{{\overline{\lambda}}^{{c\dot\alpha}}}_{u} \right) \right. 
	\nonumber \\ 
	&+& \left. \varphi^a \left( \varepsilon^{uv}  
	{\overline{\lambda}}^b_{{\dot \alpha}u} 
    {{\overline{\lambda}}^{c\dot \alpha}}_{v} 
	+ \varepsilon_{{\dot u}{\dot v}} {\psi^{b\alpha}}^{{\dot u}}  
	{\psi_{\alpha}}^{c{\dot v}} \right)  \right. \nonumber \\ 
	&+&  \left.{\overline \varphi}^a \left( \varepsilon_{u v} 
	{\lambda^{b\alpha}}^{u} {\lambda_{\alpha}}^{cv} +  
	\varepsilon^{{\dot u}{\dot v}}  
	{{\overline{\psi}}^b_{\dot \alpha}}_{{\dot u}} 
	{{\overline{\psi}}^{c\dot\alpha}}_{{\dot v}} \right) \right\}\, ,
\label{yukawa} 
\end{eqnarray} 
where $u,v=1,2$ are $SU(2)$ indices for the defining representation.

\section {Instanton Contributions to Yang--Mills 
 Correlation  Functions}  
 
We will now consider a selection of current correlation functions  in 
the $\calN=4$ Yang--Mills theory   which receive contributions 
in the one-instanton semi-classical approximation. 
Here and in the following we will restrict 
to the case of a $SU(2)$ gauge group although we will make general
comments about the  $SU(N_{c})$ case in the concluding section  
(section 7).
The special feature shared  by the particular  
observables that we will consider is that they receive a non 
vanishing contribution from the $2N_{c} \cdot {\cal N}=16$ 
gaugino zero modes that arise in the instanton background. 

We will begin by studying four-point correlation 
functions  which are expected to correspond (according to
\cite{maldacena}) to 
 non-perturbative IIB interactions that do 
not violate the ten-dimensional $U_B(1)$ symmetry of the
two-derivative part of the effective action. 
We  will then consider examples  of  correlation functions associated 
with products of composite operators with 
fields at the same point and no derivatives
(contact terms).  These should correspond to supergravity amplitudes  
that 
violate the ten-dimensional 
$U_B(1)$ symmetry but are consistent with $SL(2,\Z)$ invariance.   
 
\subsection{Correlations of four Yang--Mills supercurrents} 
 
An obvious example of a correlation function of four superconformal 
currents 
is the correlation function  
of four stress tensors. However, due to its complicated 
tensorial structure even the free-field expression for this correlator
is awkward to express compactly and it is much simpler to consider 
 correlations of four gauge-invariant  composite scalar operators, 
such as the non-singlet scalars, 
\be
\label{qdef}
{\cal Q}^{ij}=\varphi^i\varphi^j-\delta^{ij} 
\varphi_k\varphi^k/6
\ee
($i,j,k=1,2,\dots ,6$), where ${\cal Q}^{ij}$ 
is the  lowest  component of the composite  twisted chiral current 
superfield \cite{howest} ${\cal W}_{(2)}^{ij}$ of equation 
(\ref{supcur}). 
After calculating   correlation functions  of these 
currents  one can derive those of any other 
currents in 
the ${\cal N}=4$ supercurrent multiplet by making use of the  
superconformal 
symmetry \cite{howest}. A way to explicitly do it may be to resort 
to analytic superspace, recalling that 
\be {\cal Q}^{ij}(x)  \sim \left. 
{\cal W}_{(2)}^{ij}(\Upsilon) \right|_{\theta=\bar\theta=0}  \;, 
\label{npoint} 
\ee
  where $\Upsilon$ are the supercoordinates of analytic superspace  
\cite{superspace}.  Therefore, by computing correlators of ${\cal 
Q}^{ij}(x)$ and  
substituting $x\rightarrow \Upsilon$ any of the other correlation 
functions  
can in principle be obtained  by expanding in the  fermionic as well 
as in the auxiliary bosonic coordinates of analytic superspace. 
 
We will therefore begin by considering  
\be 
 \langle {\cal Q}^{i_1j_1}(x_1)  {\cal Q}^{i_2j_2}(x_2)  
 {\cal Q}^{i_3j_3}(x_3) {\cal Q}^{i_4j_4}(x_4)\rangle .
\label{qfourpoint} 
\ee
The value of this  correlation function in the  free field theory is
determined from the expression for the  free two-point scalar 
Green function which is
\be 
\la \varphi^{ia}(x) \varphi^{jb}(y)\ra_{_{\mbox{\tiny{free}}}} =  
\frac{1}{(2 \pi)^2} \frac{\delta^{ij}\delta^{ab}}{(x-y)^2}. 
\label{twopoint} 
\ee   
Therefore, the free-field expression for the correlation function  
that follows by  Wick contractions is   
\ba && \la {\cal Q}^{i_1j_1}(x_1)  {\cal Q}^{i_2j_2}(x_2) 
{\cal Q}^{i_3j_3}(x_3) {\cal Q}^{i_4j_4}(x_4)
\ra_{_{\mbox{\tiny{free}}}}  \\  \rule{0pt}{26pt}
&& \hspace{-0.1cm} = \frac{1}{(4 \pi^2)^{4}} \left[ {N_{c}}^{2} 
{\delta^{i_1 i_3}\delta^{j_1 j_3 }\delta^{i_2 i_4}\delta^{j_2 j_4} 
\over x_{13}^4 x_{24}^4 } + N_{c} 
{\delta^{j_4 i_1}\delta^{j_1 i_3}\delta^{j_3 i_2}\delta^{j_2 i_4} 
\over x_{41}^2 x_{13}^2 x_{32}^2 x_{24}^2 } + \mbox{ permutations}  
\right] \nonumber \, , 
\label{wick} 
\ea   
where, 
\be 
 x_{ij}=x_i-x_j \: . 
\label{xijdef} 
\ee  
The first term in this expression is simply the product of two  
two-point 
functions and is known to be exact.  The second term,  which  is the 
free-field  
contribution to the connected four-point function, certainly  gets 
corrections 
from interactions.   These   come from planar Feynman diagrams in the 
't~Hooft 
limit, in which $N_c\to \infty$ and  
$\hat g^2 = g_{_{YM}}^2 N_{c}$ is held fixed \cite{thooft},  
and are of  order $N_{c}$ and 
arbitrary order in $\hat g^2$.    Since $\alpha'/L^2$ is assumed to 
be  small 
the \AdS5s5/Yang--Mills correspondence requires $\hat g^2$ be large 
(see equation (\ref{dict})) and such 
interaction contributions are not under control from the point of 
view of the 
perturbative Yang--Mills theory. 

It is easy to check that in the one-instanton  background 
the correlation function (\ref{qfourpoint})  soaks up the 
sixteen  gaugino zero-modes as  
follows from the form of the supersymmetry transformations in 
(\ref{transs}).  The standard  
instanton solution  gives for the self-dual non-abelian 
field strength the expression   
\be 
 F^{a-}_{(0)\mu\nu} =  - \frac{4}{g_{_{YM}}} 
{\eta_{\mu\nu}^a \rho_0^2 \over \left(\rho_0^2  
+ (x-x_{0})^2 \right)^2}  \: , 
\label{instform}
\ee  
where $\eta^a_{\mu\nu}$ is the 't Hooft symbol \cite{thooftb}  and 
the overall power of $g_{_{YM}}$ is consistent with the normalization 
of the action (\ref{action}).  
The expression (\ref{instform}) is annihilated by the conserved  
supersymmetry transformations (those associated with the parameter 
${\overline \eta}^{\dot \alpha}_{\ A}$ in (\ref{transs})), while the 
transformations corresponding to supersymmetries 
associated with $\eta_{\alpha}^{\ A}$ are broken and generate 
eight of the sixteen fermionic zero modes. 
 
In order to evaluate this  instanton contribution  
in the most convenient fashion it will be convenient to use  the 
${\cal N}=2$ supersymmetric description in which as we explained 
before $\varphi^i$ decomposes into 
the complex $\calN=2$  singlet $\varphi$ with $U(1)$ charge $+2$ 
  and a neutral $\calN=2$ quaternion $q_{_T}$ in the  $({\bf 2,2})_0$
representation  which resides in the ${\cal N}=2$ hypermultiplet.  
It will then be sufficient to consider  the correlation function 
of two  $\varphi^2$ and two ${\overline{\varphi}}^2$ and use 
$SU(4)$ symmetry to deduce all the other components. From 
(\ref{wick})  the  free-field expression for this 
particular correlator is 
\be 
\la\varphi^2(x_1){\varphi}^2(x_2){\overline {\varphi}}^2(x_3) 
{\overline{\varphi}}^2(x_4)
\ra_{_{\mbox{\tiny{free}}}}  = \frac{1}{(4 \pi^2)^{4}}  
\left({2 N_{c}^{2} \over 
x_{13}^4 x_{24}^4} +  {N_{c} \over x_{41}^2 x_{13}^2 x_{32}^2  
x_{24}^2} 
\right) \: , 
\label{corrpert} 
\ee  
where in the previous notation $\varphi^{2} = {\cal Q}_{(+)} \equiv 
{\cal Q}^{55} - {\cal Q}^{66} + 2i {\cal Q}^{56}$.
 
In considering the one-instanton contribution to this correlation 
function one must integrate over the 
fermionic zero modes that are present in the  instanton background. 
The $\calN=2$ formalism is particularly transparent because  
it follows from the structure of the  Yukawa couplings (\ref{yukawa}) 
that $\varphi$ 
only absorbs those zero modes of the ${\cal N}=4$
 gauginos that belong to the 
${\cal N}=2$ vector multiplet ($\lambda^u_\alpha$) while    
$\overline{\varphi}$ absorbs  the  zero modes  belonging to  
the ${\cal N}=2$ hypermultiplet ($\psi^{\dot u}_\alpha$).  
We use the expressions for the zero-modes suggested by the 
supersymmetry transformations of the instanton 
\be
\label{susyzero}
\lambda_{(0) \alpha}^u =  {1\over 2} 
F_{(0)\mu\nu}^- \sigma_\alpha^{\mu\nu\, \beta} 
{1 \over \sqrt{\rho_0}} \left( \rho_0\eta_\beta^u + 
(x-x_0)_\kappa \sigma^\kappa_{\beta\dot \beta }
\bar\xi^{\dot \beta u} \right) \, .
\ee
and similarly for $\psi^{\dot u}_{(0)\alpha}$.
In equation (\ref{susyzero}) $\eta$'s and ${\overline \xi}$'s are  
constant 
adimensional spinors. In this decomposition $\eta_{\beta}$ and 
$x_{\beta{\dot \beta}}{\overline \xi}^{{\dot \beta}}$
are the parameters of the broken 
supersymmetry and special supersymmetry transformations respectively.
To recover the natural normalization in which the supersymmetry 
parameters 
have dimension (length)${}^{1/2}$ we have introduced the appropriate 
factors of
$\rho_0$. For future use notice that one can assemble the fermionic 
collective
coordinates into the spinors $\zeta^{u}_{\pm}(\rho_0,x)$ (where $\pm$ 
refers to
the $U(1)$  R-symmetry charge) to  parametrize the fermionic zero 
modes, 
\be 
\zeta^u_{\pm \alpha} (\rho_0,x-x_0) = \frac{1}{\sqrt{\rho_0}} 
\left( \rho_0 \eta^u_{\pm \alpha} +  
{\sigma^{\mu}}_{\alpha{\dot \alpha}} (x_{\mu} - x_{0\, \mu})  
{\overline \xi}^{{\dot \alpha\, u}}_{\pm} \right)
\label{zetpmdef} \:  .
\ee
%
%\nonumber \\
%\zeta^{\dot u}_{- \alpha} (\rho_0,x-x_0)& = & \frac{1}{\sqrt{\rho_0}} 
%\left( \rho_0 \eta^{\dot u}_{- \alpha} +  
%{\sigma^{\mu}}_{\alpha{\dot \alpha}} (x_{\mu} - x_{0\, \mu})  
%{\overline \xi}^{{\dot \alpha\, {\dot u} }}_{-} \right)
%
It will prove relevant in the $AdS$/Yang--Mills correspondence that  
the spinor
$\zeta$ can be rewritten as a chiral projection of a five dimensional 
spinor, 
\be 
\label{zetdef}
 \zeta (\rho_0,x-x_0) = \left( \frac{1 - \gamma^{5}}{2} \right) 
  { z^{M}  \gamma_{M} \zeta^{(0)} \over \sqrt \rho_0} \; , 
\ee  
where
\be
\label{zetzdef} 
	\zeta^{(0)} = \left( 
	\begin{array}{c} 
	    {\eta}_{\alpha} \\  \\ {\overline \xi}^{{\dot \alpha}}  
	\end{array} 
	\right) \;, \hspace{0,5cm} 
	z^{M} = (x^{\mu} - x_0^\mu,\rho_0) \; , 
\ee 
and $\gamma^{M}=(\gamma^{\mu},\gamma^{5})$ can be interpreted as 
the five dimensional Dirac 
matrices in the $AdS_5$ space (and we have suppressed the 
$SU(2)_{\cal V} \times SU(2)_{\cal H} \times U(1)$ R-symmetry 
indices $u$ and $\dot{u}$).    
In fact, as we will see later, these spinors turn out to be chiral 
projections of the  Killing spinors of $AdS_5$ which is crucial for 
the correspondence with D-instanton calculations.  

With the above normalization conventions, the jacobian of
the fermionic part of the  instanton measure is (in the $N_c=2$ case 
at hand)
\be
J_F  = \left( {g_{_{YM}}^2 \over 2^5 \sqrt{2} \pi^2 \rho_0} \right)^8 
\,  .
\label{fermjac}
\ee
One can also compute the jacobian for the transformation from 
bosonic
zero-modes  to the collective coordinates that leads to the familiar 
bosonic
measure for $SU(N_c)$ instantons \cite{bernard}
\be
\label{measdef}
d\mu_{B(K=1)} = {2^3 \over \pi^2} {(2\pi^2)^{2 N_c} \over 
(N_c-1)!(N_c-2)!}\left({\rho_0 \over g_{_{YM}}}\right)^{4N_{c}} 
{d\rho_0 d^4x_0 \over \rho_0^5} \: ,
\ee
where we will only make use of the case $N_c =2$.\footnote{The 
combined 
(fermionic and bosonic) subtraction point dependence, 
$\mu^{(4-{\cal N}) N_{c}}$, disappears for 
${\cal N}=4$ as well as for all other superconformal theories.} 

Using the preceding  equations one finds that in the 
semi-classical approximation
\cite{thooftb,akmrv} the one-instanton contribution to the four 
bosonic current Green function is\footnote{Here and in the following 
computations of correlators of currents bilinear in the
fundamental Yang--Mills fields a factor of $g^2_{_{YM}}$ is included 
for each external
insertion.  
It would be equivalent to rescale the fundamental Yang--Mills fields
according to $A_{_{YM}} \to A'_{_{YM}}=g_{_{YM}} A_{_{YM}}$ 
and compute the Green functions of primed fields. This 
leads to a common overall dependence on $g^2_{_{YM}}$ for the three 
correlation functions that we will consider.}
\ba  
&& G_{{\cal Q}^4}(x_{p}) = \la  g_{_{YM}}^2 \varphi^2(x_1) 
g_{_{YM}}^2  \varphi^2(x_2) g_{_{YM}}^2 \overline{\varphi}^2(x_3) 
 g_{_{YM}}^2 \overline{\varphi}^2(x_4) \ra_{_{K =1}} = {g_{_{YM}}^8  
\over 32 
\pi^{10}} 
 \nonumber \\ \rule{0pt}{20pt}
&& \hspace{-0.7cm} e^{-{8\pi^2 \over g_{_{YM}}^2} + i\theta_{_{YM}}} 
\int {d\rho_0 d^4x_0 \over \rho_0^5} d^4\eta_+  
d^4\overline{\xi}_+d^4\eta_-d^4\overline{\xi}_- \:  
\varphi_{(0)}^2(x_1)\varphi_{(0)}^2(x_2) 
\overline{\varphi}_{(0)}^2(x_3)\overline{\varphi}_{(0)}^2(x_4)  , 
\nonumber\\
\label{corrinst} 
\ea where the subscript $K=1$ denotes the winding number of the 
background and  
$(+)$ (or $(-)$) refers to the $U(1)$ charges of the gauginos in the 
vector (or hyper) multiplet (see below). The Green function
$G_{{\cal Q}^4}$ receives a contribution also from the
$K=-1$ sector that is the complex conjugate of (\ref{corrinst}).

In (\ref{corrinst})  the fields $\varphi$ and 
$\overline{\varphi}$ have been  replaced  by the expressions 
\ba 
&& \varphi(x) \rightarrow \varphi_{(0)}(x) =  \frac{1}{2\sqrt{2}}
\varepsilon_{uv} \zeta^{u}_{+} \s^{\mu\nu} \zeta^{v}_{+}   
F^-_{(0)\mu\nu}  \nonumber \\ 
&& \overline{\varphi}(x) \rightarrow 
\overline{\varphi}_{(0)}(x) =  
\frac{1}{2\sqrt{2}} \varepsilon_{\dot{u}\dot{v}} \zeta^{\dot u}_{-} 
\s^{\mu\nu} \zeta^{\dot v}_{-}  
F^-_{(0)\mu\nu} \: , 
\label{scalarsol} 
\ea   
which are the leading nonvanishing terms that     
result from  Wick contractions in which Yukawa  couplings are lowered 
from the exponential of the action until a sufficient number of 
fermion fields are present to saturate the fermionic integrals.    
Of course, these expressions can also be obtained 
directly from the supersymmetry transformations (\ref{transs}) 
by acting twice on $F^{-}_{(0)\mu\nu}$ with the broken supersymmetry 
generators. 
After some elementary Fierz transformations on the fermionic  
collective 
coordinates the fermionic integrations can be performed in a standard 
manner and the result is 
\be  
G_{{\cal Q}^4}(x_{p}) = { 3^4   \over 2^{27} \pi^{10}}\, g_{_{YM}}^8
\,e^{-{8\pi^2 \over g_{_{YM}}^2} + i\theta_{_{YM}}} 
\,   \int {d\rho_0 d^4 x_{0} \over \rho_0^5} \: 
x_{12}^4 x_{34}^4 \, \prod_{p=1}^4 {\left( {\rho_0 \over 
{\rho_0^2 + (x_p - x_{0})^2}}  \right)}^4 . 
\label{collectint} 
\ee
As mentioned earlier, the fact that  the  instanton form factor,
$\rho_0^4/[(\rho_0^2 + (x_p - x_0)^2]^4$, that enters this expression 
is identical to  
$ K_4$ in (\ref{greenfun})  will be of significance in the
discussion of the $AdS$/Yang--Mills correspondence in
section 5.    
The integration  in (\ref{collectint})
resembles that of a standard Feynman diagram with  
momenta 
replaced by position differences and can be performed by introducing 
the Feynman parametrization, 
\ba  
G_{{\cal Q}^4}(x_p) &=& {3^4 \Gamma(16) \over  2^{27} \pi^{10} \left(
 \Gamma(4) 
 \right)^{4} }\, g_{_{YM}}^8   \, 
e^{-{8\pi^2 \over g_{_{YM}}^2} + i\theta_{_{YM}}} \,   
\int \prod_{p} {\alpha}_p^3 d {\alpha}_p \delta 
\left( 1-\sum_q \alpha_q \right) \nonumber \\ 
&& \hspace{-0.5cm}  \int {d\rho_0 d^4 x_{0} \over \rho_0^5}   
{x_{12}^4 x_{34}^4 \: \rho_0^{16} \over   (\rho_0^2 + 
x_{0}^2 - 2 x_{0} \cdot \sum_p \alpha_p x_p + \sum_p x_p^2)^{16}} \: 
. 
\label{feynman} 
\ea   
The five-dimensional integral yields 
\ba  
G_{{\cal Q}^4}(x_p) &=& \frac{3^{3} \, \Gamma(11)}
{\left( 2^8 \pi^3 \Gamma(4) \right)^{4}} \:
 g^{8}_{_{YM}}e^{-{8\pi^2 \over g_{_{YM}}^2} + i\theta_{_{YM}}}   
\nonumber \\
&& \hspace{-0.7cm}  \int \prod_p \alpha_p^3 d\alpha_p \, 
\delta \left(1-{\sum}_q  \alpha_q \right) 
{x_{12}^4 x_{34}^4 \over 
(\sum_p \alpha_p \alpha_q x_{pq}^2 )^{8}} \: . 
\label{integral} 
\ea   
This integral can be simplified by observing that it is essentially  
obtained by acting with derivatives on the box-integral with four  
massless external particles, 
\be  
G_{{\cal Q}^4}(x_p) =\frac{3^{3} \, \Gamma(11)}
{\left( 2^8 \pi^3 \Gamma(4) \right)^{4}} \:
  g^{8}_{_{YM}} e^{-{8\pi^2 \over g_{_{YM}}^2} + i\theta_{_{YM}}}  \,
x_{12}^4 x_{34}^4 \prod_{p<q} 
{\partial \over \partial x_{pq}^2} B(x_{pq}) \, , 
\label{intbox} 
\ee   
where the box integral is 
\be B(x_{pq}) = \int {\prod_{p} d\alpha_p  
\delta \left( 1-\sum_q \alpha_q \right)  
\over (\sum_p \alpha_p \alpha_q x_{pq}^2 )^{2}} \: . 
\label{boxint} 
\ee  
 
The result may be expressed as  a combination of 
dilogarithms\footnote{We thank Lance Dixon 
for pointing out a sign error 
in the version of this formula in \cite{dix} and suggesting where 
to find the correct version in \cite{bern}.}, 
\ba  
B(x_{pq}) &=& {1\over \sqrt{\Delta(x_{pq})}} \left[ 
- {1\over 2} \log\left ({u_+u_-\over(1-u_+)^2(1-u_-)^2}\right) 
\log\left({u_+\over u_-} \right) \right.  \nonumber  \\  
&& \hspace{-2.3cm} - \left. \mbox{ Li}_2(1-u_+) + \mbox{ 
Li}_2(1-u_-)  
-\mbox{ Li}_2\left( 1-{1\over u_-} \right) + 
\mbox{ Li}_2\left( 1-{1\over u_+} \right) \right] 
\label{boxeplicit} 
\ea   
where,   
\be 
\Delta(x_{pq}) = \det_{4\times 4}((x^2_{pq})) =    
X^2 + Y^2 + Z^2 - 2 XY - 2 YZ - 2 ZX, 
\label{det} 
\ee   
and 
\be 
u_{\pm} = { Y + X - Z \pm \sqrt{\Delta} \over 2 Y} \: , 
\label{ratio} 
\ee   
with $X = x_{12}^2 x_{34}^2$, $Y = x_{13}^2 x_{24}^2$ and  
$Z=x_{14}^2 x_{23}^2$.

Notice that up to an overall dimensional factor needed for the correct
scaling, $G_{{\cal Q}^4}$ turns out to be a function of the two
independent superconformally invariant cross ratios  $X/Z$ and $Y/Z$.
 Although not immediately apparent, the expression $B(x_{pq})$
is   symmetric under any permutation of the external legs, 
as can be seen by making use of the properties  of the dilogarithms, 
\ba  
\mbox{ Li}_2(z) + \mbox{ Li}_2 (1-z) &=& {\pi^2\over 6} - \log(z) 
\log(1-z) \cr   
\mbox{ Li}_2(z) + \mbox{ Li}_2 \left({1\over z} \right) &=& 
-{\pi^2\over 6} - [\log(-z)]^2 \: , 
\label{dilog} 
\ea  
and observing that the relevant permutations correspond to  
permutations of $X= Y u_+u_-$, $Y$   and $Z= Y(1-u_+)(1-u_-)$, 
that are generated by the two 
transformations: a) $u_+\rightarrow 1/u_-$, $u_-\rightarrow 1/u_+$,  
$Y\rightarrow Y u_- u_+$, which is equivalent to the exchange of $X$ 
and $Y$, 
leaving $Z$ fixed  (\ie to the exchange of $x_1$ and $x_4$ or, 
equivalently, of 
$x_2$ and $x_3$) and  b) $u_+\rightarrow 1-u_-$, $u_-\rightarrow 
1-u_+$ at fixed $Y$, which is  equivalent to the 
exchange of $X$ and $Z$ (or the exchange of $x_2$ and $x_4$ or, 
equivalently,  of  $x_1$ and $x_3$). 

Unlike  correlation functions of elementary 
fields that are  infra-red problematic and gauge-dependent, the above 
correlator is well defined at non-coincident points.  
Up to the derivatives acting on the box integral, the result is 
exactly the one expected for the correlator of four scalar operators  
of dimension  $\Delta = 2$ each. The detailed expression 
for this 
correlator, including the $g^8_{_{YM}}$ factor,  will later be 
related to  the single D-instanton contribution 
to  the ${\cal R}^4$ term in the type 
IIB  effective action around the $AdS$ background.

\subsection{ The correlation function of  sixteen fermionic currents} 
 
As could have been anticipated, 
it is particularly simple to analyze the 
contribution of the Yang--Mills instanton to the correlation function 
of sixteen of the fermionic superconformal current bilinears, 
${\hat \Lambda}_\alpha^{A} = \Tr ( \sigma^{\mu 
\nu}{}_{\alpha}{}^{\beta} 
F^{-}_{\mu \nu}  {\lambda_{\beta}}^{A})$, 
\begin{equation} 
G_{\hat \Lambda^{16}}(x_p) = 	\langle \prod_{p=1}^{16} g_{_{YM}}^2 
\hat \Lambda^{A_{p}}_{\alpha_{p}}(x_{p})  
	\rangle_{_{K=1}} \; , 
	\label{l16}
\end{equation} 
Since each factor of $\hat\Lambda$ in the product 
provides a single fermion zero mode it is necessary to consider the 
product of sixteen currents  in order  to saturate the sixteen  
grassmannian integrals. To leading order in $g_{_{YM}}$, 
$G_{\hat \Lambda^{16}}$ does not receive contribution from anti-instantons.
The leading term in the one-instanton sector 
is simply obtained by replacing  each $F^{-}_{\mu \nu}$ with the  
instanton profile  ${F^{-}_{(0)\mu \nu}}$ 
(equation (\ref{instform})) and each  
$\lambda_\alpha^{\ A}$ with the corresponding 
zero mode, ${\lambda_{(0)\alpha}}^{A}$, 
that can be deduced from the action of the broken supersymmetry in 
the second line of (\ref{transs}),
\be
\label{lamzero}
\lambda_{(0) \alpha}^A =  {1\over 2} 
F_{(0)\mu\nu}^- \sigma_\alpha^{\mu\nu\, \beta} 
{1 \over \sqrt{\rho_0}} \left( \rho_0\eta_\beta^A + 
(x-x_0)_\mu \sigma^\mu_{\beta\dot \beta }
\bar\xi^{\dot \beta A} \right) \, .
\ee
The resulting correlation function thus has the form 
\begin{eqnarray} 
	G_{\hat \Lambda^{16}}(x_p) 
    &=& { 2^{11} 3^{16} \over \pi^{10} }\, 
  g_{_{YM}}^{8} e^{-{8\pi^2 \over g_{_{YM}}^2} + i\theta_{_{YM}}} 
    \int \frac{d^{4}x_{0} \, d\rho_0}{\rho_0^{5}}  
	\int d^{8}\eta d^{8}{\overline \xi}  \nonumber \\  
	&& \hspace{-2.6cm}  \prod_{p=1}^{16} \left[ 
	\frac{\rho_{0}^{4}}{[\rho_0^{2}+(x_{p}-x_{0})^{2}]^{4}} \,
	\frac{1}{\sqrt{\rho_{0}}}
	\left( \rho_0 \eta^{A_{p}}_{\alpha_{p}}+
	{(x_{p}-x_{0})}_{\mu}  \sigma^{\mu}_{\alpha_{p}{\dot \alpha}_{p}}  
	{\overline \xi}^{{\dot \alpha}_{p}A_{p}} \right)  \right] \: . 
	\label{l16-sym} 
\end{eqnarray} 
The integration over the fermion zero modes leads to a 
 sixteen-index invariant  tensor, $t_{16}$, of the product of the  
$SU(4)$ and Lorentz groups. Assembling the 16 fermionic collective
coordinate into a sixteen-dimensional spinor $t_{16}$ would simply read
$t_{16}^{a_1a_2\dots a_{16}} = \varepsilon^{a_1a_2\dots a_{16}}$, with
 $a_i = 1,2 \dots 16$.
 Further integration over 
the instanton moduli space would determine the   
dependence on the coordinates $x_{p}$.  However, we will leave the 
expression in the unintegrated form (\ref{l16-sym})  which will be 
compared later with the corresponding expression obtained in the IIB 
string theory in \AdS5s5.  Again,  the resemblance of the instanton
 form factor to the $AdS$ Green function will be of significance to
 the discussion in section 5.

\subsection{ The  correlation function of eight gaugino 
bilinear currents}

In similar fashion it is easy to deduce the one-instanton 
contribution to other related processes, such as the eight-point 
correlation function, 
\be  
G_{{\calE}^8}(x_p) = \langle g_{_{YM}}^{2} {\cal E}^{A_1B_1}(x_1) 
\dots 
g_{_{YM}}^{2}{\cal E}^{A_8B_8}(x_8) \rangle_{_{K=1}}.  
\label{eeightpoint} 
\ee   
which also saturates the sixteen fermionic zero-modes present 
in the $SU(2)$ one-instanton background. To leading order in 
$g_{_{YM}}$, 
$G_{{\calE}^8}$ does not receive contribution from anti-instantons.
The complete non-abelian expression for ${\cal E}^{AB}$ is 
\be 
{\cal E}^{AB} = \lambda^{\alpha a A} {\lambda_{\alpha a}}^{B} + 
g_{_{YM}} f_{abc} \, t_{ijk}^{(AB)_+} \phi^{ia}\phi^{jb}\phi^{kc} 
\label{exacte} 
\ee 
but  at leading order in the gauge coupling constant only the term  
proportional to the gaugino bilinear is relevant.  
In the instanton background the gaugino 
bilinear is given by 
\be  
\lambda_{(0)}^{\alpha a A} {\lambda_{(0)\alpha a}}^{B} =  
{3\cdot 2^6 \over g_{_{YM}}^2} 
\frac{\rho_0^{4}}{\left( \rho_0^{2}+(x-x_{0})^{2}\right)^{4}}
\: \zeta^{\alpha A} {\zeta_{\alpha}}^{B} \, . 
\label{bilin} 
\ee 
It follows that 
\ba  
    \hspace{-0.5cm} &&G_{{\calE}^8}(x_p)
    = {3^8 2^{11} \over \pi^{10}} \, 
    g_{_{YM}}^{8} \, e^{-{8\pi^2 \over g_{_{YM}}^2} + 
i\theta_{_{YM}}} 
    \int \frac{d^{4}x_{0} \, d\rho_0}{\rho_0^{5}}\int d^{8}\eta
      d^{8}{\overline \xi}  \nonumber \\ 
     && 
	  \prod_{p=1}^{8} \left[ 
	\frac{\rho_0^{4}}{\left( \rho_0^{2}+(x_{p}-x_{0})^{2}\right)^{4}}
	\frac{1}{\sqrt{\rho_0}}
	\left( \rho_0 \eta^{A_{p}}_{\alpha_{p}}+
	{(x_{p}-x_{0})}_{\mu}  
	\sigma^{\mu}_{\alpha_{p}{\dot \alpha}_{p}}  
	{\overline \xi}^{A_p {\dot \alpha}_p}\right)  
	\right. \  \nonumber \\ 
	\hspace{-0.5cm} &&  \left. 
	\varepsilon^{\alpha_p \beta_p} \frac{1}{\sqrt{\rho_0}}
	\left( \rho_0 \eta^{B_{p}}_{\beta_{p}}+
	{(x_{p}-x_{0})}_{\nu}  
	\sigma^{\nu}_{\beta_{p}{\dot \beta}_{p}}  
	{\overline \xi}^{ B_p {\dot \beta}_p}\right)\right]  
  \label{e8-sym} 
\ea 
The integration over the fermion zero modes leads to an 
$SU(4)$ invariant contraction of
sixteen-index tensor $t_{16}$ defined after (\ref{l16-sym})
and further integration over 
the instanton moduli space would determine the   
exact dependence on the coordinates $x_{p}$.  Again the unintegrated 
expression (\ref{e8-sym}) is sufficient for comparison with the 
D-instanton contribution to the corresponding \AdS5s5\ amplitude.

\section{D-instanton effects in $D=10$ IIB theory} 
  
We now turn to consider the single D-instanton contribution to the  
amplitudes 
in the IIB superstring theory that are related to the  above 
Yang--Mills 
correlation  functions according to the correspondence suggested in 
\cite{maldacena}.   
 
The two leading terms in the momentum expansion of the IIB effective 
action that involve only the metric and the complex scalar fields are 
given (in string frame) by \cite{greengut}, 
\ba
&&(\alpha')^{-4} \int d^{10}X \sqrt g \left(e^{-2\phi} R + k 
(\alpha^\prime)^3
 f_4(\tau,\bar\tau) 
e^{-\phi/2} {\cal R}^4    \right) \nonumber\\
&& =  L^{-8} \int d^{10}X \sqrt g \left((4\pi N_c)^2 R + k L^6 (4\pi 
N_c)^{1/2}
 f_4(\tau,\bar\tau)  {\cal R}^4    \right),
\label{twoterms}
\ea
where the relations (\ref{dict}) have been substituted in the
second line
and $k= 1/(2^{11} \pi^7)$, as follows, for instance, by direct 
comparison of 
the result of \cite{greengut} with equation (15) of \cite{gkt}.
As discussed in outline in \cite{banksgreen} the $f_4(\tau,\bar\tau)
{\cal R}^4$ interaction
contains the D-instanton terms that should be compared with the 
effects of Yang--Mills instantons.
The Riemann curvature enters the ${\cal R}^4$ factor in a manner 
that    
may be most compactly described by writing it as an integral over a   
sixteen-component Grassmann spinor, 
\be 
{\cal R}^4 \equiv \int d^{16}\Theta (R_{\Theta^4})^4, 
\label{rfourw}  
\ee  
where  
\be\label{rthet} R_{\Theta^4} = \overline{\Theta} \Gamma^{\Lambda_1 
\Lambda_2  
\Lambda} 
 \Theta \, \overline{\Theta} \Gamma^{\Lambda_3\Lambda_4}_{\ \ \ \ \ 
\Lambda} 
 \Theta \, R_{\Lambda_1\Lambda_2\Lambda_3\Lambda_4}, 
\ee which only includes the Weyl tensor piece of the Riemann 
tensor.   
Here, $\Gamma^{\Lambda_1 
\Lambda_2  
\Lambda_3}$ are the totally antisymmetric products of three 
ten-dimensional
$\Gamma$-matrices and the 
Grassmann parameter $\Theta^a$ ($a=1,\dots,16$) is a  chiral spinor 
of the 
ten-dimensional theory.   This expresses ${\cal R}^4$ as an integral
over half of the  
on-shell type IIB superspace, which accounts for the fact that this 
term has 
exactly determined properties.  In \cite{banksgreen} it was pointed
out that the  
${\cal R}^4$  term vanishes in the \AdS5s5\ background 
because it involves a fourth power of the (vanishing) Weyl tensor.
The first, second and third functional derivatives of ${\cal R}^4$
vanish as well, and as a result one finds no corrections to zero, one,
two and three-point amplitudes.
However, there is a non-zero four-graviton
amplitude  
arising from this term. The boundary values of these gravitons are
sources for various bosonic components of the Yang--Mills current 
supermultiplet.   For example, the components of the metric  in the
$AdS_5$ directions couple to the stress tensor, ${\cal T}^{\mu\nu}$,
whereas the traceless components polarized in the $S^5$ directions couple to  
massive Kaluza-Klein states. A linear combination of the trace of the
metric on $S^5$ and the fluctuation of the \RR\ four-form potential
couples to the scalar components, ${\cal Q}^{ij}$.

It is easy to pick out 
 all of the other terms of the same dimension that are related to 
 ${\cal R}^4$ by supersymmetry by, for example, associating the
 physical fields with the components of an on-shell IIB superfield 
\cite{howesttwo}.  Included among these is a sixteen-fermion 
interaction \cite{greengutkwon},
\be
\label{lambint} 
(\alpha')^{-1} \int d^{10}X \sqrt g \, e^{-\phi/2} 
f_{16}(\tau,\overline{\tau})  \Lambda^{16}   + c.c.,
\ee
 where $\Lambda$ is a complex chiral $SO(9,1)$ spinor which transforms
 under the $U_B(1)$ R-symmetry with charge 3/2 and the 
 interaction is antisymmetric in the sixteen spinor indices. 
 The function $f_{16}$ is related to $f_4$ by 
\be  
f_{16} = (\tau_{2} {\cal D})^{12} f_4, 
\label{relf} 
\ee  
where ${\cal D}=(i\partial/\partial \tau - 2q/\tau_{2})$  
is the covariant  derivative 
that maps a $(q,p)$ modular form into a $(q+2,p)$ form   
(where the notation 
$(q,p)$ labels the holomorphic and  anti-holomorphic $SL(2,\Z)$
weights of the form).  
Whereas $f_4$ transforms with modular weight $(0,0)$ 
the function $f_{16}$ has 
weight $(12,-12)$  and therefore transforms with a specific
phase under $SL(2,\Z)$.   This is precisely the phase required to
compensate for the $U_B(1)$ transformation of the $\Lambda^{16}$ 
factor 
so that the full expression (\ref{lambint}) is invariant under 
$SL(2,\Z)$.

All of these terms can be expanded in the limit of small 
 coupling (large $e^{-\phi}$) in the form 
\be 
f_n = a_n \zeta(3)e^{-3\phi/2}  +  
  b_n e^{\phi/2} + \sum_{K =1}^\infty \calG_{K,n} e^{\phi/2}, 
\label{fnexp} 
\ee  
where the first two terms (with constant coefficients $a_n$ and $b_n$)
have the form of string tree-level 
 and one-loop terms and $\calG_{K ,n}$ contains  the charge-$K$ 
 instanton and anti-instanton terms.  
The instanton contribution to ${\cal G}_{K,n}$ has  the asymptotic 
expansion in powers of $e^{\phi}$, 
\be 
\calG_{K ,n} =  \mu(K) (K e^{-\phi})^{n -7/2} e^{-2\pi K (e^{-\phi} 
+  i  C^{(0)}) } \left(1 + \sum_{k=1}^\infty c_{k,n}^K (K 
e^{-\phi})^{-k}
\right) \: , 
\label{gmdef} 
\ee  
while the anti-instanton contributions will not be considered here.
The coefficients $c_{k,n}^K$ are explicitly given in  
\cite{greengutkwon} and 
\be\label{zmeas} \mu(K) = \sum_{m|K} {1\over m^2}, 
\ee where $m|K$ denotes the sum  over all divisors of $K$.  The 
expression (\ref{gmdef}) has the appropriate factor of $e^{-2K \pi 
e^{-\phi}}$ to be associated with a charge-$K$ D-instanton effect.   
The complete expansion in expression (\ref{gmdef}) represents a 
series of perturbative  fluctuations   
around a D-instanton  where the exact coefficients  depend on  the 
precise form of the interaction that involves $n$ type IIB fields.  
However, the leading term  in (\ref{gmdef}) can be written as 
\be 
\calG_{K ,n} \sim \calZ_K (K e^{-\phi})^n   , 
\label{zndef} 
\ee  
where  ${\cal Z}_K$ is independent of which particular interaction 
term  is 
being discussed and is given by 
\be 
\calZ_K = \mu(K) (K e^{-\phi})^{-7/2} e^{-2\pi K (e^{-\phi} +  
i C^{(0)}) } \: , 
\label{znew} 
\ee  
${\cal Z}$ should be identified with the contribution of a charge-$K$  
D-instanton to the measure in string frame which,  
up to an overall numerical factor $c$, we write
\be
\label{dmeasdef}
d \mu_{K}^{(s)} = c (\alpha')^{-1}
d^{10} X \, d^{16}\,  \Theta\,
{\cal Z}_K.
\ee
In doing this we are being
cavalier about the fact that the full series 
in (\ref{gmdef}) is not convergent (it is actually an asymptotic
approximation to a Bessel function).  In the end, consistency of the 
full theory, particularly with modular invariance, should require 
considering the complete expression for the instanton contribution. 
  
As observed in \cite{banksgreen}\ the charge-$K$  
D-instanton action that appears in the exponent in (\ref{znew}) 
coincides with the action of a charge-$K$ Yang--Mills instanton in 
the 
boundary theory which indicates a correspondence between 
 these sources of non-perturbative 
effects.    This idea  is reinforced by the correspondence 
between other  factors.   For example, after substituting 
$e^\phi = g_{_{YM}}^2/4\pi$ and 
$\alpha' = L^2 N^{-1/2} g^{-1}_{_{YM}}$, the measure 
(\ref{dmeasdef}) contains an overall factor of the coupling constant 
in the 
form $g_{_{YM}}^8$.
Indeed, this is exactly the power expected on the
 basis of the  
\AdS5s5/SYM\  correspondence   since, as we have seen, 
the one-instanton contribution to the Green functions
 in the ${\cal N}=4$ Yang--Mills theory, considered in section 3,
also has a factor of 
$g_{_{YM}}^8$ arising from the combination  of the bosonic and 
fermionic zero 
modes norms.   We will pursue this issue further in the next section 
by comparing  the leading 
instanton contributions  to  IIB superstring amplitudes with the  
corresponding  $\calN=4$ current correlators considered in section 
3, even though a  complete matching would be surprising in 
$N_c=2$ case under consideration.

\section{$AdS$/Yang--Mills correspondence}

We will now compare the one-instanton contribution to the 
supersymmetric 
Yang--Mills correlation functions presented in section 3 with the 
amplitudes of
the IIB   superstring theory with appropriate boundary conditions. In 
the case of the bulk IIB theory the D-instanton
effects may be either extracted directly, as in this section, from 
the exactly 
known terms in the effective action or deduced, as in the next 
section, from the integration of the semi-classical fluctuations 
around the 
$AdS_5\times S^5$ D-instanton solution.

The  \AdS5s5\  IIB background is endowed with a metric which may be 
defined in terms of  ten-dimensional 
Cartesian coordinates $(x^\mu, y^i)$  ($i= 1,\cdots,6$) by, 
\be  
ds^2 =  {L^2 \over \rho^2} (dx\cdot dx + dy \cdot dy ) =  
{L^2 \over \rho^2}(dx\cdot dx  + d\rho^2) + d{\omega_5}^2  ,
\label{cfmetric} 
\ee 
where  $\rho^2 =  y^2$ and $d\omega_5^2$ 
is the spherically-symmetric  constant curvature metric on $S^5$.
The \AdS5s5 background is characterized by the non-vanishing fields, 
\ba 
F_{_{MNPQR}} &=& \frac{1}{L} \varepsilon_{_{MNPQR}}   
\qquad R_{_{MNPQ}} = - \frac{1}{L^2} (g_{_{MP}}g_{_{NQ}} - 
g_{_{MQ}}g_{_{NP}})\\ 
F_{mnpqr} &=& \frac{1}{L} \varepsilon_{mnpqr} 
\qquad\quad R_{mnpq} = + \frac{1}{L^2} (g_{mp}g_{nq} - g_{mq}g_{np}) 
\label{adsxs} 
\ea 
(recall that upper case Latin indices, $M,N,\dots=0,1,2,3,5$, span the
$AdS_5$ coordinates and lower case Latin indices, $m,n,\dots 
=1,2,3,4,5$
span the $S^5$ coordinates).  
 The only non-vanishing components of the Ricci tensor are 
\be  
R_{MN} = - \frac{4}{L^2} g_{MN} \qquad 
R_{mn} = + \frac{4}{L^2} g_{mn} \, .
\label{ricci} 
\ee 
Upon contracting (\ref{ricci}) with the  metric tensor it follows 
that  the 
scalar curvature vanishes.  

This background is maximally supersymmetric (just like the
Minkowski vacuum) so there are 32 conserved 
supercharges that transform as a
complex chiral spinor of the tangent-space group, $SO(4,1) \times
SO(5)$.
In the basis where the ten dimensional $\Gamma_\Lambda$ matrices are 
given by 
$\Gamma_M = \s_1\otimes\gamma_M\otimes \I$ and
$\Gamma_m=\s_2\otimes \I\otimes\gamma_m$, the supersymmetries are 
generated by  
the Killing spinors that satisfy  
\be  
D_\Lambda \epsilon - {1 \over 2L}(\s_1\otimes \I\otimes\I) 
\Gamma_\Lambda
\epsilon = 0 ,
\label{kilspina} 
\ee 
which follows from the requirement that the  gravitino 
supersymmetry transformation should vanish. 
In this basis the complex chiral supersymmetry parameter reads
\ba
\epsilon=\left(\begin{array}{c}1\\0\end{array}\right)
\otimes \zeta \otimes \kappa \, , 
\label{spindecomp}
\ea
where $\zeta$ is a
complex four-component $SO(4,1)$ spinor  and $\kappa$ a complex
four-component $SO(5)$ spinor. The Killing spinor equation 
(\ref{kilspina})
has components   
\ba  
D_M \zeta - {1 \over 2L} \gamma_M \zeta &=& 0 
\label{kilspon}
\\ 
D_m \kappa - i {1 \over 2L} \gamma_m \kappa &=&0  ,
\label{kilspin} 
\ea 
The Killing spinors have a number of interesting properties.
In particular, the Killing spinors $\kappa$ on $S^5$ may
be  used to construct the Kaluza-Klein excitations of all the fields 
in 
the IIB 
gauged supergravity starting from the modes of the  massless  complex
singlet  dilaton $e^\phi$.   For future purposes, 
it is convenient to also consider the spinors that satisfy
the Killing spinor equation with the opposite relative sign between 
the two 
terms in (\ref{kilspon}) and (\ref{kilspin}).
We will denote the former by  $\zeta_+$ and $\kappa_+$ and the latter 
by $\zeta_-$ and 
$\kappa_-$.  In the next section, we will also need the Euclidean 
continuation of the Killing spinors on $AdS_5$.

The identification of the lowest lying modes that form the 
supergravity supermultiplet is given in \cite{kiromvan}  and is rather
involved.   Indeed, while the massless dilaton is associated
with the constant   mode on  
$S^5$, \ie with the scalar spherical harmonic $Y_{\ell}$ with 
$\ell=0$, 
the other scalars in the supermultiplet are associated with 
excitations on the 5-sphere. 
In particular the real scalars $Q^{ij}$ with mass $m^2 = -4/L^2$ in 
the {\bf 20}$_{_{\tiny{\R}}}$ of the 
$SO(6)$ isometry group of $S^5$ result from a combination of the trace
of the internal metric and the self-dual \RR\ five-form field,
$F^{(5)}$, with $\ell=2$ ($Q^{ij}$ are  quadrupole moments of $S^5$).  
Similarly the complex scalars $E^{AB}$ with mass $m^2 = -3/L^2$ and 
their 
conjugates are associated with the pure two-form fluctuations with
$\ell=1$ of the  
complexified antisymmetric tensor in the internal directions. 
The {\bf 15} massless vectors $V_M^{[ij]}$ that gauge the $SO(6)$
isometry group  
are in one-to-one correspondence with the Killing vectors of $S^5$ and
result  
from a linear combination with $\ell=1$ of the mixed components of the
metric  
and the internal three-form components of the \RR\ four-form
potential, 
$C^{(4)}$. The {\bf 6} complex  
antisymmetric tensors $B_{MN}^{[AB]}$ with $m^2=1/L^2$, that have 
peculiar 
first order 
equations of motion, result from scalar spherical harmonics with 
$\ell=1$. The analysis of the fermions is similar. 
The {\bf 4} dilatini $\Lambda^A$ with mass $m = - 3/(2L)$ are 
proportional to the internal Killing spinors $\kappa_+$. The {\bf 
20}$_{_{\tiny{\sC}}}$  
spinors $\chi_{BC}^A$ 
with mass $m=-1/(2L)$ correspond to internal components of the
gravitino  
with 
$\ell=1$. Finally the supergravity multiplet is completed by the 
massless {\bf 4}$^*$ gravitinos $\Psi_{MA}$ which  are 
proportional to the internal Killing spinors $\kappa_-$.  

The above fields are the ones   that act as sources for the 
superconformal
currents (\ref{currentdef}). 
Higher Kaluza--Klein  modes have higher values of  $\ell$.   For
example, there are
scalar modes like $Q^{ij}$ with $\ell>2$.  Each of these
can be put in one-to-one correspondence 
with a gauge singlet composite operator $W_{(\ell)}$ that starts with 
\be 
\left. W^{(i_1...i_\ell)}\right|_{\theta=0} = \Tr \, \phi^{(i_1}
...\phi^{i_\ell)}  
- \mbox{ traces}  
\label{kkwell} 
\ee 
which has dimension 
$\Delta=E_0 = \ell$ and belongs to the $\ell$-fold symmetric 
traceless 
tensor representation of $SO(6)$. The multiplet contains 
$256\ell^2(\ell^2-1)/12$ 
states with different $E_0$ and $SU(4)$ quantum numbers. 
 
It is important to recall 
 that the non-compact $E_{6(+6)}$ symmetry of the ungauged IIB 
theory  
that results from a compactification on $T^5$ is broken by the gauging
 of the  
$SO(6)$ subgroup of $E_{6(+6)}$ leaving an unbroken $SL(2,\R)$ which is
identified with the original global 
 symmetry of the ten-dimensional theory.
The classical gauged supergravity lagrangian possesses a local $Sp(8)$
 symmetry inherited from the ungauged theory under which  
the  fermions transform as {\bf 8} and the scalar vielbein as a {\bf 
27} 
(antisymmetric traceless tensor). But minimizing the scalar potential 
induced by  gauging results in a spontaneous breaking of $Sp(8)$ to
 $U(4)$.  
The $U(1)$ factor may be thought as a remnant of the local $U_B(1)$
 symmetry of the flat ten-dimensional classical supergravity 
theory, but there is no
 corresponding symmetry  in the $D=4$ Yang--Mills boundary theory.
 This fits in 
  with the fact that in IIB string theory  the $SL(2,\R)$ is replaced
 by  $SL(2,\Z)$ and the continuous  $U_B(1)$ symmetry is not  
present.    
For example, the presence of the $\Lambda^{16}$ interaction in the IIB
 effective action is consistent with
 $SL(2,\Z)$ but not with the $SL(2,\R)$ symmetry of the minimal 
classical IIB supergravity.

We may now  compare the effects of Yang--Mills instantons in
supercurrent correlators evaluated in section 3 
with those of D-instantons in the \AdS5s5\
IIB string theory.   For illustrative purposes the simplest choice is 
to
compare the D-instanton contribution to the  amplitude  for  sixteen 
spin-1/2 
gluinos, $\Lambda$,  in the \AdS5s5\ background   with the Yang--Mills
instanton contribution to the sixteen-${\hat \Lambda}$ correlation
function.  From either perspective 
the leading instanton contribution arises from the product of sixteen 
factors each carrying one single fermionic zero mode.

The first method for obtaining this amplitude is to 
expand the function $f_{16}$ in (\ref{relf}) to extract the
one-instanton term.  In order to compare with the Yang--Mills
sixteen-point correlator we need to consider the situation in which
all sixteen fermions propagate to well defined configurations
on the boundary.  
The Dirac operator acting on spin-1/2 fields in $AdS_{5}$ was given 
in \cite{henningsfet} by  
\begin{equation} 
	\ms{D}\Lambda  = {e_{{\hat L}}}^{M} \gamma^{\hat{L}}  
	\left( \partial_{M} +\frac{1}{4} 
	\omega^{\hat{M}\hat{N}}_{M} \gamma_{\hat{M}\hat{N}}
	\right)\Lambda
 =  	(\rho  \gamma^{\hat 5}  
	\partial_{5}  
	+ \rho \gamma^{\hat\mu} \partial_{\mu} - 2 \gamma^{\hat 5})
\Lambda \; , 
	\label{dirac-op} 
\end{equation} 
where ${e_{\hat{L}}}^{M}$ is the vielbein,
$\omega^{\hat{M}\hat{N}}_{M}$  the  
spin connection (hatted indices refer to the tangent
space) and $\gamma^{ \hat\mu}$ are the  four-dimensional Dirac 
matrices.  
Equation (\ref{dirac-op}) leads to the normalized 
bulk-to-boundary propagator of the fermionic field $\Lambda$ 
of mass $m=-3/2L$, associated to the composite operator  
${\hat \Lambda}$ of dimension $\Delta = \frac{7}{2}$, 
\begin{equation} 
K^F_{7/2}(\rho_0,x_0;x) = K_{4}(\rho_0,x_{0};x) {1 \over 
\sqrt{\rho_0}}
\left( \rho_0 \gamma_{\hat5} +  {(x_{0}-x)}^{\mu}
	\gamma_{\hat\mu}	\right)
	\label{eq:prop-7/2} ,
\end{equation}   
which, suppressing all spinor indices, leads to  
\be
 \Lambda_{J}(x_0,\rho_0) = \int d^4 x K^F_{7/2}(\rho_0,x_{0};x) 
J_{\Lambda}(x) \, ,
\label{boundarylambda}
\ee
where $J_{\Lambda}(x)$ is a left-handed  boundary value of 
$\Lambda$ and acts as the 
source for  the composite operator  
${\hat \Lambda}$  in the boundary $\calN=4$ Yang--Mills theory.
As a result, the classical action for the operator ${(\Lambda)}^{16}$
in the   
$AdS_{5}\times S^{5}$ supergravity action is
\begin{eqnarray}  
  S_{\Lambda} [J] &=& e^{-2\pi ({1\over g_s}  +  
  i  C^{(0)})} g_s^{-12}  
  \, V_{S^5} \, \int \frac{d^{4}x_{0}d\rho_0}{\rho_0^{5}} \nonumber\\
  && \hspace{-0.8cm} t_{16} \prod_{p=1}^{16}  
    \left[ K_{4}(\rho_0,x_{0};x_{p}) \frac{1}{\sqrt{\rho_0}}
	\left( \rho_0 \gamma^{\hat 5} +  
    {(x_{0}-x_{p})}^{\mu} \gamma_{\hat\mu} \right)  
    \, J_{\Lambda}(x_{p}) \right] \, ,
	\label{source-l16} 
\end{eqnarray} 
where we have set $e^\phi = g_s$ and $C^{(0)} = \tilde C^{(0)}$ 
(since the scalar fields are taken to be 
constant in the \AdS5s5\ background) and $V_{S^5}=\pi^3$ 
is the $S^5$ volume. The 16-index invariant tensor $t_{16}$ is the same
as the one defined after (\ref{l16-sym}). 
The overall power of the coupling constant comes from the factor of 
$g_s^4$ in the measure (\ref{dmeasdef}) and the factor of
 $g_s^{-16}$ from the leading term in (\ref{gmdef}) (and we have 
dropped an overall numerical constant).
Using the dictionary (\ref{dict}) and 
 differentiating with respect to the chiral sources  
this result agrees with the  
expression (\ref{l16-sym}) obtained in the Yang Mills  
calculation, including the power of $g_{_{YM}}$ (but we have not 
checked the overall numerical constant). 
In the next section we will also motivate this expression directly by 
semi-classical quantization around a D-instanton field configuration 
that is a Euclidean solution of the IIB supergravity.

The agreement of the $\Lambda^{16}$ amplitude  with the corresponding 
sixteen-current correlation function of section 3 is sufficient to 
guarantee that the instanton contributions to all the other 
Yang--Mills  correlation functions that are related to this term  
by ${\cal N}=4$ supersymmetry will also agree with their IIB 
superstring counterparts.  For example, 
the correlation function that we considered in most detail
 in section 3 was the one with four superconformal scalar currents 
${\cal Q}^{ij}$ which are in the {\bf 20}$_{_{\tiny{\R}}}$ 
of $SU(4)$, and have  dimension $\Delta=2$ and $AdS$ mass 
$m^2=-4/L^2$. 
The supergravity field, $Q^{ij}$, that couples to ${\cal Q}^{ij}$ is  
a linear 
combination of the fluctuation of the trace of
the metric on $S^5$, $h_{m}^{\ m}$,  and of the four-form field
potential, 
$C^{(4)}_{MNPQ}= \epsilon_{MNPQR} \nabla^Rf$.
Therefore, contributions to the correlation of four of these 
composite scalars in the one D-instanton background should 
correspond to the leading parts of the  
$K=1$ terms in   the expansion of the ${\cal R}^4$  
interaction (\ref{rfourw}) as well as terms  of the form ${\cal R}^2
(\nabla F_5)^2$ and $(\nabla F_5)^4$.  These last two terms involve
$F_5$, the
self-dual  field strength of the antisymmetric four-form potential, 
and   are related  by supersymmetry to the ${\cal R}^4$ term.

It follows from the structure of  
(\ref{collectint}) that the Yang--Mills  instanton contribution 
to each factor of ${\cal Q}^{ij}$ is of the form   
${\nabla\nabla} K_2(x_0^\mu,\rho_0 ;x^\mu,0)$, where the two 
derivatives are
not necessarily contracted.   
 But this is the  expected form for a
 propagator  from the \AdS5s5\ bulk to the 
boundary for a scalar field of dimension 2.   
Therefore, at least the general form of the   expression obtained  
from the 
 $K=1$ terms in the expansion of ${\cal R}^4$ and the related
$F_5$ interactions agrees with  
the four-${\cal Q}^{ij}$ correlation function in a Yang--Mills 
instanton background. 
 In order to see this agreement 
in more detail it would first be necessary to determine the 
precise form of the $(\nabla F_5)^2$ and $(\nabla F_5)^4$  
interactions that  
contain the fluctuations of $F_5$.   
The $K=1$ contribution  to the amplitude of four fluctuations of the 
appropriate combination of $h_{m}^{\ m}$ and $f$  can then  be 
extracted  from these interaction terms.
Although we have not so far performed this explicitly, the result is
guaranteed to reproduce the $K=1$ expression obtained from the
Yang--Mills theory in section 3 since it is related by supersymmetry
to the $\Lambda^{16}$ amplitude.  

The analogous comparison of the  correlation  function of eight  
${\cal E}^{AB}$'s 
with the amplitude for eight $E^{AB}$'s  in the IIB
superstring theory proceeds in much the same way. The supergravity 
fields,
$E^{AB}$, that couple to ${\cal E}^{AB}$ arise from the internal 
components
of the complex IIB antisymmetric tensor. Supersymmetry relates an
$H^8$ term, where $H$ is the complex IIB three-form field-strength, 
to the ${\cal R}^4$ term in the IIB effective action. One thus expects
interactions schematically of the form $(\nabla K_3)^8$, with various
contractions of the derivatives. Using the explicit form of the spinor
collective coordinates $\zeta^A=(\rho_0\eta^A +
(x-x_0)_\mu\s^\mu\bar\xi^A)/\sqrt{\rho_0}$ one may check that the 
gaugino
bilinears ${\cal E}^{AB}$ in the one-instanton background exactly 
give rise to
$K_3$ and derivatives thereof, as  expected for the scalar propagator 
of a field
of dimension
$\Delta=3$ and $AdS$ mass $m^2=-1/L^2$. Although
a precise matching of the resulting amplitudes is still
under study, one may appeal to supersymmetry arguments to determine 
the  complete structure of these terms.

Notice that the above considered non-perturbative terms in the IIB 
effective
action when expanded around the 
$AdS_5\times S^5$ give rise to both derivative and non-derivative
interaction terms.  
The matching of the non-derivative ``mass-related'' terms with 
corresponding
terms in the Yang--Mills Green functions is rather trivial but 
clearly it is
only a hint to the conjectured correspondence.

\section{The D-instanton solution in \AdS5s5}

We will now  consider to what extent the  information about the 
charge-one D-instanton term   extracted from the 
$(\alpha')^{-1}$ terms  
in the IIB effective action can be determined by semi-classical 
IIB supergravity field theory in a D-instanton background. 
Recall that in flat ten-dimensional euclidean space the charge-$K$  
D-instanton solution is a finite-action euclidean 
supersymmetric (BPS--saturated) solution  in which
the  metric is trivial ($g_{\mu\nu} = \eta_{\mu\nu}$ 
in the Einstein frame) but the complex scalar $\tau = C^{(0)} + i 
e^{-\phi}$ has a nontrivial profile with a singularity at the 
position of the D-instanton.   The (euclidean) 
\RR\ scalar is related to the dilaton  by the BPS condition  
$\partial_\Sigma C^{(0)} = \pm i 
\partial_\Sigma e^{-\phi}$, while the dilaton solution is 
the harmonic function (correcting an error in \cite{ggp}) 
\be 
e^{\hat \phi^{(10)}}  = g_s + { 3 K {\ap}^4 \over \pi^4 |X-X_0|^8}. 
\label{tendim} 
\ee
This is the classical solution  of the ten-dimensional 
Laplace equation,
 $\partial^2e^\phi =0$,  outside  an
infinitesimal sphere centered on the point $X_0^\Lambda$ (where 
 $X^\Lambda$ is the ten-dimensional coordinate 
and $X_0^\Lambda$ is the location of the D-instanton), 
$g_s$ is the asymptotic value of the string coupling and the 
normalization of the second term has a quantized value by virtue of a 
condition analogous to the Dirac--Nepomechie--Teitelboim condition 
that quantizes the charge of an electrically charged   
$p$-brane and of its magnetically charged $p'$-brane dual 
\cite{nepo,teit}.  
It is notable that the solution in (\ref{tendim}) is simply the Green 
function for a scalar field to propagate from $X_0$ to $X$ subject to 
the boundary condition that $e^\phi = g_s$ at $|X| \to \infty$ or 
$|X_0| \to \infty$.

We are now  interested in solving the equations of motion of the IIB 
theory in euclidean  \AdS5s5. 
The   BPS condition for a D-instanton in this background again  
requires $\partial_\Sigma  e^{-\phi}  = \pm i \partial_\Sigma C^{(0)}$ that 
leads to  
\be 
g^{\Lambda\Sigma} \nabla_\Lambda \nabla_\Sigma e^\phi = 0,
\label{disteq} 
\ee 
and (in the Einstein frame) the Einstein equations are unaltered by
the presence of the D-instanton (because the associated Euclidean stress 
energy tensor vanishes) so  that \AdS5s5\ remains a solution.
Equation (\ref{disteq}) is  
identical to  the equation for the Green function of a 
massless scalar propagating between the location of the D-instanton
($x_0^\mu, y_0^i$) and the point ($x^\mu,y^i$), which is the
bulk-to-bulk propagator (subject to the  boundary condition that it is
constant in the limits $\rho\to 0$ and  $\rho \to \infty$).
This is easy to solve
using the conformal flatness of \AdS5s5\ which implies that the
solution for the dilaton  
is of the form\footnote{We are grateful to G.W.  Gibbons and M.J. 
Perry for
discussions about the general form of this solution.}
\be
\label{conflat}
e^{\hat \phi}  = g_s +  { \rho_0^4 \rho^4\over L^8} 
\left(e^{\hat \phi^{(10)}} - g_s \right),
\ee
where  $\rho_0 = |y_0|$ and $e^{\hat \phi^{(10)}}$ 
is the harmonic function that appeared in the flat ten-dimensional 
case,
(\ref{tendim}).\footnote{This does {\it not} 
agree with the expressions recently
proposed in \cite{chuhowu} or in \cite{kogan}
that appeared while this paper was in  preparation.}
In evaluating D-instanton dominated  amplitudes 
we will only be interested in the case in which the point
$(x^\mu,y^i$) approaches the boundary ($\rho \equiv |y| 
\to 0$), in which case it is necessary to rescale the 
dilaton profile (just as it is necessary to rescale the scalar
bulk-to-bulk propagator, \cite{gkp,wittone}) so that the combination
\be
\rho^{-4} \left(e^{\hat \phi} - g_s \right) =   
{3 K(\alpha')^4 \over L^8 \pi^4}{  \rho_0^4\over  ((x-x_0)^2 + 
\rho_0^2)^4 }  \, , 
\label{classd}
\ee
will be of relevance in the $\rho \to 0$ limit.     

As mentioned earlier,
the correspondence with the Yang--Mills instanton 
follows from the fact that $  \rho_0^4/ ((x-x_0)^2 + 
\rho_0^2)^4  = K_4$
 is proportional to the instanton number  density, $(F_{(0)}^-)^2$, 
in the $\calN=4$
 Yang--Mills theory.  Strikingly, the  scale size of the Yang--Mills
 instanton is replaced by the distance $\rho_0$ of the D-instanton 
from
 the boundary. This is another indication of how the geometry of the
 Yang--Mills theory is encoded in the IIB superstring.   Note,
 in particular,  that 
as the D-instanton approaches the boundary $\rho_0 \to 0$, 
the expression for  $\rho^{-4} e^{\hat \phi}$ 
reduces to a $\delta$ function
 that corresponds to a zero-size Yang--Mills instanton.

The BPS condition implies that we can write the solution for the \RR\
scalar as
\be
\label{rrhat}
\hat C^{(0)} = \tilde C^{(0)} + i f(x,y),
\ee
where $\tilde C^{(0)}$ is the constant real part of the field (which
corresponds to $\theta_{_{YM}}/2\pi$ see equation (\ref{idenscal})) and 
\be
\label{fsoln}
f =A -  {1\over g_s} + e^{-\hat \phi}.
\ee 
Since the action is independent of constant shifts of $C^{(0)}$ it does not 
depend on the arbitrary constant, $A$. 
In a manner that follows closely  the flat ten-dimensional case 
considered in the appendix of \cite{greengut}  the  action for a single 
D-instanton of charge $K$ can be written as 
\be
\label{actint}
S_K = - {L^{10}\over (\alpha')^4} \int  
{d\rho d^4x d^5 \omega \over \rho^5}  g^{\Lambda\Sigma} 
\nabla_\Lambda (e^{2\hat \phi} f \partial_\Sigma f),
\ee
which reduces to an integral over the boundaries of \AdS5s5\ and the surface 
of an infinitesimal sphere centered on the D-instanton at $x=x_0$, $y=y_0$.  
With the choice $A=0$ in (\ref{fsoln}) the 
entire D-instanton action comes from the boundary of 
the infinitesimal sphere.  Substituting for $f$ from (\ref{fsoln}) gives 
\be 
S_K = {2\pi |K|\over g_s},
\label{dinstact} 
\ee 
which is the same answer as in the flat ten-dimensional case.  
On the other hand,  with the choice $A=1/g_s$  in (\ref{fsoln})  the  
expression (\ref{actint}) reduces to an integral over the  boundary 
at $\rho=0$  but the
total action remains the same as $S_K$ in (\ref{dinstact}).   
Remarkably,  in this case the boundary integrand  is {\it identical} 
to the action density  of the standard four-dimensional Yang--Mills instanton. 

Whereas the \AdS5s5\ metric remains unchanged by the presence of the
D-instanton in the Einstein frame it is radically altered in the
string frame where the instanton is manifested as a space-time
wormhole (as in the flat ten-dimensional case \cite{ggp}).  For finite
values of $K$  the 
dilaton becomes large in the Planck-scale neck and the classical solution 
is not reliable  in that region.   However, for very large
instanton number, the neck region becomes much larger than the Planck
scale so, by analogy with the D-brane examples studied in
\cite{maldacena}, it should be very interesting to study the
implications of the modified \AdS5s5\ geometry in the large-$K$ limit
of the large-$N_c$ theory.

The D-instanton contribution to the amplitude with
sixteen external dilatinos, $\Lambda_{\alpha}^A$, may now be 
obtained directly by
semi-classical quantization around the classical
D-instanton solution in \AdS5s5. 
  The leading instanton contribution can be determined by applying
supersymmetry transformations to the scalar field  which has an 
instanton profile given by  (\ref{classd}).   Since 
the D-instanton background breaks half the supersymmetries the
relevant transformations are those in which the supersymmetry
parameter corresponds to the  Killing  
spinors for the sixteen broken supersymmetries.  
These Killing spinors have  $U_B(1)$ charge 1/2  and are  defined by a 
modified version of (\ref{kilspina})
that includes   the non-trivial composite $U_B(1)$ connection, $Q_M$
\cite{shw},  that is made from  the IIB scalar field \cite{ggp}, 
\be
\label{modsixty}
{\cal D}_M \zeta \equiv (D_M - {i\over 2} Q_M)\zeta  = {1 \over 2L} 
\gamma_M \zeta
\ee  
Substituting the euclidean D-instanton solution into the expression
for the composite connection gives 
\be
\label{Qdef}
Q_M =  {i\over 2} e^{-\hat \phi} \partial_M e^{ \hat \phi} 
\ee
with $\hat \phi$ defined
by (\ref{classd}).  The solution of (\ref{modsixty}) is
\be 
\zeta_{\pm} = 
e^{-\hat\phi/4} {z_{M} \gamma^{\hat M} \over \sqrt{\rho_0}} 
\zeta^{(0)}_{\pm}, 
\label{spinkil} 
\ee 
where $\zeta^{(0)}_{\pm}$ is a constant spinor satisfying 
$\gamma_5 \zeta^{(0)}_{\pm} = \pm \zeta^{(0)}_{\pm}$.  
 
The sixteen broken supersymmetry transformations associated
with $\zeta^{(0)}_{-}$  give rise to the  dilatino
zero-modes,
\be 
\Lambda_{(0)} = \delta \Lambda = (\gamma^M \hat P_M) \zeta_{-}, 
\label{dil5d} 
\ee
where $\hat P_M$ is the expression for $P_M \equiv i\partial_M 
\tau^*/2\tau_2$  in the D-instanton background
\cite{greengut},
\be
\label{pmdef} \hat P_M =   e^{-\hat \phi} \partial_M  e^{\hat \phi}.
\ee 
Using the Killing spinor equation and the D-instanton equation
${\cal D}^M \hat P_M=0$  
it is easy to check (recalling that $P_M$ has $U_B(1)$ charge 2) that 
\be
\label{lambdeq}
\gamma^M {\cal D}_M \Lambda_{(0)} = - {3 
\over 2L}  
\Lambda_{(0)},
\ee
 so that $\Lambda_{(0)}$ is a solution of the
appropriate massive Dirac equation.  
We will be interested in amplitudes with external states located on the  
boundary, in which case  we may use the fact that for $\rho \sim 0$,
\be
\label{farapp}
 P_M \sim{1\over  g_s} \partial_M e^{\hat \phi}
\ee
 in (\ref{dil5d}),  which  leads to 
\be\label{lamzer}
\Lambda_{(0)} \sim  {4\rho^4\over g_s} (e^{\hat \phi} -  g_s) \zeta_{-}
 \, .  
\ee
This means that near $\rho=0$   the dilatino profile in the D-instanton 
background   is proportional to $\rho^{4} K_4(x_0,\rho_0;x, 0)$.  

As a result the leading contribution to the sixteen-dilatino  
amplitude again reproduces the corresponding sixteen-current 
correlator in ${\cal N}=4$  supersymmetric 
Yang--Mills theory. 
Explicitly,  the D-instanton approximation to the  amplitude with  
sixteen external dilatinos, ${\Lambda_\alpha}^{A}$, at points on the 
$\rho=0$ 
boundary is (up to an overall constant factor)
\begin{eqnarray} 
	\langle \prod_{p=1}^{16} 
	\Lambda^{A_{p}}_{\alpha_{p}}(x_{p},0) \rangle_J 
    &=& g_s^{-12} e^{-2\pi K ({1\over g_s} +  i  C^{(0)}) } 
    V_{S^5} \,  \int \frac{d^4x_0 d\rho_0}{\rho_0^{5}}  
	\int d^{16}\zeta^{(0)}_{-} 
	 \nonumber \\  
	&& \hspace{-3.8cm}  \prod_{p=1}^{16} \left[ K_{4} 
	\left(x_0,\rho_{0};x_{p}\right) {1\over\sqrt{\rho_0}}   
	\left(\rho_0 \eta^{A_{p}}_{\alpha_{p}}+{(x_{p}-x_{0})}_{\mu}  
	\sigma^{\mu}_{\alpha_{p}{\dot \alpha}_{p}}  
	{\overline \xi}^{{\dot \alpha}_{p}A_{p}} \right)
 J_{\Lambda}(x_p) \right],  
	\label{l16dil5d} 
\end{eqnarray} 
where $J_{\Lambda}(x_p)$ is the wave-function of the dilatino
evaluated at the boundary  point $(x_p,0)$ and the Grassmann spinor 
$\zeta_-^{(0)}$ was defined in terms of $\eta$ and $\bar \xi$ in 
(\ref{zetzdef}).  The power of $g_s^{-12}$ 
has been inserted in (\ref{l16dil5d}) from the expression obtained in 
section 4 although this power should also follow directly by 
considering the normalization of the bosonic and fermionic zero modes.
Up to the overall constant factor, the  amplitude (\ref{l16dil5d})  
 agrees with (\ref{source-l16}) and therefore with
(\ref{l16-sym}).   
 
In similar manner the instanton profiles of all the fields in the
supergravity multiplet follow by applying the broken supersymmetries
to $P_M$ any number of times, just as they do in the flat
ten-dimensional case \cite{greengut}.  The single D-instanton
contributions to any correlation function can then be determined.
This is guaranteed to agree with the corresponding term in the
expansion of the effective type IIB action as well as with
the corresponding ${\cal N}=4$ Yang--Mills correlation function. 

\section{Discussion} 
 
We have analyzed instanton contributions to correlation functions of 
superconformal currents in $\calN=4$ Yang--Mills theory in three 
ways.  
The first was a direct Yang--Mills instanton calculation at lowest 
order in perturbation theory around the one-instanton configuration. 
The superconformal correlation functions that we considered are ones 
in 
which the sixteen gaugino zero modes implied by supersymmetry in the 
instanton 
background are soaked up by the currents.  We considered the 
particular example of  the four-${\cal Q}$ correlation function in 
most detail, reducing it to an explicit conformal invariant 
expression involving dilogarithms.  We also obtained integral
representations for the sixteen-$\hat \Lambda$ and eight-${\cal E}$ 
correlation functions.   These correlation functions and many others 
are related to each other by   the superconformal symmetry.

The Yang--Mills calculations  were then 
compared with single D-instanton contributions to 
corresponding amplitudes in the \AdS5s5\ compactification of the IIB 
superstring.   These contributions were isolated from the exactly 
known form 
of the appropriate interaction terms in the IIB effective action.  
The sixteen-$\Lambda$ amplitude was considered in detail since it is 
the simplest to compare with the corresponding Yang--Mills 
expression.  

Finally,  we derived the classical D-instanton solution in the \AdS5s5\ 
background.  Quite strikingly we saw that the D-instanton action could be 
expressed as a surface integral in two equivalent ways.  In one of these the 
action is localized at the position of the D-instanton in the ten-dimensional 
euclidean space, $(x_0,y_0)$, while in the  other the action is given by an 
integral over the $AdS$ boundary at $\rho=0$.
In the latter case the integrand is precisely the same as the action density 
of a Yang--Mills instanton in four euclidean dimensions.
We then saw that the sixteen-$\Lambda$ amplitude obtained 
by semi-classical quantization  agreed with the exact expression (although 
we have not determined the precise constant factor in the measure).

The fact that the Yang--Mills instanton and IIB D-instanton effects 
appear to match so closely should be interpreted  as general support for the  
conjectured   correspondence of
\cite{maldacena}.  Notably we have seen that the Yang--Mills 
instanton scale size has a natural interpretation as the position of 
the D-instanton in extra dimensions transverse to the 
four-dimensional boundary space-time. This reflects the fact that the 
single-instanton moduli spaces in both cases contain an $AdS_5$ factor.
The  additional dimensions, which  are not at all apparent  in  standard
${\cal N}=4$ Yang--Mills  perturbation theory, are very natural
when the  \AdS5s5\  is interpreted as the near-horizon geometry of
$N_c$  D3-branes
as in \cite{maldacena}.  The  points in the space transverse to
the boundary, with coordinates $y^i$, are then identified 
with the expectation values  of the six massless 
scalar fields in a separated test D3-brane and $\rho^2=|y|^2$. 
The five  angular coordinates of $S^5$,
$\omega$, are  the additional  massless scalar fields that complete the
moduli space to $AdS_5 \times S^5$.

  However, the instanton effects were obtained at leading order in 
$g_s = 4\pi g^2_{_{YM}}$ and next-to-leading order in $(g^2_{_{YM}} 
N_c)^{-1/2}= \alpha'L^{-2}$  and  should therefore only match 
precisely for gauge group $SU(N_c)$ in the limit of large-$N_c$.  
Since we have
only considered the case $N_c=2$ there are many  obvious questions 
and  it is
likely that the agreement we have found  can only be made precise by 
extending
the calculations in several  directions.  
Firstly, the exact expressions for the IIB interactions include an 
infinite number of fluctuations around the D-instanton and it would 
be an interesting project in its own right to study the corresponding 
series of fluctuations around the Yang--Mills instanton.
Secondly, it would be interesting to consider larger $SU(N_c)$ gauge 
groups.  This requires a careful 
discussion of zero modes.  
For example, consider the single instanton contribution  when 
$N_c>2$.   
The number of bosonic collective coordinates is known to be $4N_c$, but 
only five of these correspond to translations and dilatation whereas 
the 
others are associated with    global $SU(2)$ or coset rotations of the 
instanton inside 
$SU(N_c)$.  For correlators of gauge-invariant operators the latter 
integrations are trivial and   one   ends up with a five-dimensional 
integral over  the moduli  of the $SU(2)$ one-instanton  
configurations and this coincides with $AdS_5$. Since there are
$8N_c$ fermionic zero modes in the instanton background it might
superficially appear that the correlation functions considered earlier
would vanish.  But this neglects the fact that all but 16 of these
are effectively eliminated by the effects of the Yukawa,  
gauge and scalar self couplings of the fields.  We have so far used 
the fact 
that   the fundamental  fields, $A$, $\lambda$, $\phi$ and 
$\overline\lambda$ absorb, respectively, $0,1,2,3$ fermionic zero
modes.   However,  when $N_c>2$ these fields have more fermionic zero
modes.  Schematically, we may write, 
\ba 
DF &=& g_{_{YM}} \left( \overline\lambda \lambda + \phi D \phi \right)
\nonumber \\ 
D\lambda &=& g_{_{YM}} \phi \overline\lambda  \nonumber \\ 
D^2 \phi &=& g_{_{YM}} \left( \lambda \lambda +  
\overline\lambda\overline\lambda \right) + g_{_{YM}}^2 \phi^3 \nonumber \\ 
D \overline\lambda &=& g_{_{YM}} \phi \lambda  \,  .
\label{eqmott} 
\ea 
Once the lowest order solution for $A$, $\lambda$, $\phi$ and 
$\overline\lambda$ 
is plugged into (\ref{eqmott}) one finds  new contributions with {\it 
four} 
more zero-modes  must be  added to each field. The process ends after 
exactly $2N_c$ cycles due to the anticommuting nature of the 
fermionic 
collective coordinates. Thus, it should be straightforward to 
generalize the $K=1$ discussion to gauge groups with  $N_c >2$.
It may considerably simplify the analysis to work in a manifestly 
supersymmetric context, such as that described 
in \cite{siegel,howest}, which allows for a fully non-linear analytic 
superspace extension of the Yang--Mills instanton configuration.

The contributions of multiply-charged ($K>1$)  instanton  
configurations  are subtle for $N_c>2$. 
The instanton moduli space, 
including the three global $SU(2)$ rotations, is a hyperk\"ahler 
quotient defined by the ADHM equations. The process of absorbing the 
fermionic 
zero-modes   appears to end up with an 
integration over a much 
larger space than   $AdS_5$.   However, one may speculate that in the 
large $N_c$  limit the $K$-instanton measure is concentrated around 
charge-$m$ $SU(2)$ instantons (where $m$ is a divisor of $K$) that 
are embedded in  $K/m$ commuting $SU(2)$ subgroups
 of $SU(N_c)$.  This   
would reduce the domain of integration to  symmetric products
of $K/m$ $AdS_5$ factors.  These spaces have orbifold singularities at 
which the number of fermionic zero modes may be reduced so that the 
special correlators we have considered may get non-zero support only 
from these regions since they correspond to local effective 
interactions.  
Clearly, it would be very interesting to  deduce the  precise 
charge-$K$
D-instanton weight,
$\mu(K)$ (given by (\ref{zmeas})),  that enters into the IIB interactions 
 by considering such Yang--Mills instanton embeddings 
in $SU(N_c)$  for  large $N_c$.  Presumably, this would shed light on 
how the moduli space of multiply-charged
D-instantons on \AdS5s5\ can be represented by
the Yang--Mills instantons beyond the most  obvious $AdS_5$ factors.

One approach to determining this measure might be to view the 
instanton-dominated
supercurrent correlators as a  
topological subset of the ${\cal N}=4$ Yang--Mills theory.  The  
modular-covariant 
non-holomorphic functions of the complexified couplings of the
D-instanton description appear to be analogous to expressions found 
in the
computation of the Witten index for topologically twisted   
${\cal N}=4$ theories on curved manifolds \cite{vafawitten}.
Other superconformally invariant gauge theories with
fewer supercharges \cite{vafetal} may similarly 
give rise to interesting topological subsectors
in which the correlation functions are dominated by instanton 
configurations
analogous to those that enter the computations performed in this 
paper.
 
In summary,  the nonperturbative
instanton effects that we have considered in this
paper seem to be very naturally adapted to
 the possibility that the dynamics of ${\cal N}=4$ Yang--Mills theory 
can be  considered to be governed by a theory living on
 the boundary of \AdS5s5\ \cite{maldacena}, although this 
correspondence is very obscure in perturbation theory. 
 Even for the gauge group $SU(2)$
 the Yang--Mills 
instanton scale is naturally identified with the distance of
 the D-instanton from the boundary. This appears to be yet more
 evidence, albeit qualitative since we have only studied the $N_c=2$
 case,  in favour of the  Maldacena conjecture.

 \vskip 0.4cm

{\it Acknowledgements}  
 
We acknowledge useful discussions with T. Banks, R.
Dijkgraaf, P. Di Vecchia, S. Ferrara,  D. Freedman, G.W. Gibbons, S. 
Gubser, 
H. Osborn, M.J. Perry, A.M. Polyakov, K. Ray, A. Sagnotti, Ya. 
Stanev, M.
Testa, and G. Veneziano. We would also like to  
acknowledge useful e-mail correspondence with L. Dixon.   Two of us
(MB, MBG) are grateful to the organizers of Strings98 at the Institute
for Theoretical Physics, Santa Barbara and particularly to John
Schwarz for providing us with office facilities.

\end{document}